\documentclass[journal]{IEEEtran}
\usepackage{url}
\usepackage{amsmath}
\usepackage{graphicx}
\usepackage{tikz}

\ifCLASSOPTIONcompsoc
    \usepackage[caption=false, font=normalsize, labelfont=sf, textfont=sf]{subfig}
\else
\usepackage[caption=false, font=footnotesize]{subfig}
\fi
\usepackage{multirow}
\usepackage{multicol}
\usepackage{adjustbox}
\usepackage{amssymb}
\usepackage{bbm}
\usepackage{booktabs}
\usepackage{float}
\usepackage{tabu}
\usepackage{bm}
\usepackage[noadjust]{cite}
\usepackage{enumitem}
\usepackage{dblfloatfix}

\usepackage{pifont}
\newcommand{\cmark}{\ding{51}}%
\newcommand{\xmark}{\ding{55}}%

\usepackage{xcolor}
\usepackage[linesnumbered,ruled,vlined]{algorithm2e}
\allowdisplaybreaks

\SetCommentSty{mycommfont}

\SetKwInput{KwInput}{Input}                
\SetKwInput{KwOutput}{Output}              
\SetKwInput{KwInit}{Init.}              
  
\definecolor{mygreen}{rgb}{0,0.6,0}
\definecolor{mygray}{rgb}{0.5,0.5,0.5}
\definecolor{mymauve}{rgb}{0.58,0,0.82}

\DeclareRobustCommand\encircle[1]{\tikz[baseline=(char.base)]{\node[shape=circle,fill=gray,inner sep=2pt] (char) {\textcolor{white}{#1}}}}

\allowdisplaybreaks

\usepackage[T1]{fontenc}

\usepackage[compact]{titlesec}
\titlespacing{\section}{0pt}{2ex}{1ex}
\titlespacing{\subsection}{0pt}{1ex}{0ex}
\titlespacing{\subsubsection}{0pt}{0.5ex}{0ex}
\usepackage{setspace} 
\begin{document}

\title{SURT 2.0: Advances in Transducer-based Multi-talker Speech Recognition}

\author{Desh Raj, \IEEEmembership{Student Member, IEEE}, Daniel Povey, \IEEEmembership{Fellow, IEEE}, and Sanjeev Khudanpur, \IEEEmembership{Member, IEEE}
\thanks{This work was partially funded by the National Science Foundation CCRI program via Grant No. 2120435, and a fellowship from Amazon via the JHU-Amazon Initiative for Interactive Artificial Intelligence (AI2AI).}
\thanks{Desh Raj is with the Center for Language and Speech Processing, Johns Hopkins University, Baltimore, MD 21218 USA (e-mail: draj@cs.jhu.edu).}
\thanks{Daniel Povey is with Xiaomi Corp., Beijing, China (e-mail: dpovey@gmail.com).}
\thanks{Sanjeev Khudanpur is with the Center for Language and Speech Processing, and Human Language Technology Center of Excellence, Johns Hopkins University, Baltimore, MD 21218 USA (e-mail: khudanpur@jhu.edu).}
}



\maketitle

\begin{abstract}
The Streaming Unmixing and Recognition Transducer (SURT) model was proposed recently as an end-to-end approach for continuous, streaming, multi-talker speech recognition (ASR).
Despite impressive results on multi-turn meetings, SURT has notable limitations:
(i) it suffers from \textit{leakage} and \textit{omission} related errors; 
(ii) it is computationally expensive, due to which it has not seen adoption in academia; and 
(iii) it has only been evaluated on synthetic mixtures. 
In this work, we propose several modifications to the original SURT which are carefully designed to fix the above limitations. 
In particular, we (i) change the unmixing module to a mask estimator that uses dual-path modeling, (ii) use a streaming zipformer encoder and a stateless decoder for the transducer, (iii) perform mixture simulation using force-aligned subsegments, (iv) pre-train the transducer on single-speaker data, (v) use auxiliary objectives in the form of masking loss and encoder CTC loss, and (vi) perform domain adaptation for far-field recognition. 
We show that our modifications allow SURT 2.0 to outperform its predecessor in terms of multi-talker ASR results, while being efficient enough to train with academic resources. 
We conduct our evaluations on 3 publicly available meeting benchmarks --- LibriCSS, AMI, and ICSI, where our best model achieves WERs of 16.9\%, 44.6\% and 32.2\%, respectively, on far-field unsegmented recordings. 
We release training recipes and pre-trained models: \texttt{\url{https://sites.google.com/view/surt2}}.
\end{abstract}


\section{Introduction}
\label{sec:intro}

The last decade has seen tremendous advancement in the transcription of single-speaker utterances~\cite{Amodei2016DS2,Xiong2017TowardHP}, resulting from the adoption of deep learning methods and large-scaled supervised training~\cite{li2021recent}.
With conversational speech recognition (ASR) systems reaching super-human performance, researchers are now turning towards more challenging settings such as multi-talker conversations~\cite{Barker2015TheT, Kinoshita2013TheRC, Watanabe2020CHiME6CT}. 
This setting is often characterized by overlapping speech, turn-taking, and far-field audio, and therefore requires special modeling techniques to tackle the problem~\cite{Carletta2005TheAM, Shriberg2001ObservationsOO, Yoshioka2019MeetingTU}. 
Despite these challenges, multi-talker ASR, particularly in the context of meeting transcription, has a long history originating from the NIST Rich Transcription evaluation~\cite{Fiscus2007TheRT,Hain2012TranscribingMW}, and its subsequent application in recognizing dinner-party conversations~\cite{Barker2015TheT,Watanabe2020CHiME6CT,Segbroeck2019DiPCoD}.

A conventional modeling approach for multi-talker ASR is through a cascade of separation and recognition systems. 
This approach leverages advancements in speech separation research to obtain single-speaker audio~\cite{Wang2017SupervisedSS}, which can then be used with a regular ASR component~\cite{Raj2020IntegrationOS}; 
however, such a model may be sub-optimal since the components are independently optimized, and may also require greater engineering efforts for maintenance~\cite{Wu2021InvestigationOP}. 

Due to these limitations with cascaded systems, researchers have proposed jointly optimized models that combine separation and ASR and directly solve for the task of multi-talker transcription, often using a permutation-invariant training (PIT) objective~\cite{Yu2017RecognizingMS}. 
Such a paradigm has been explored in the context of hybrid HMM-DNN systems~\cite{Qian2017SingleChannelMS}, and more recently for end-to-end ASR~\cite{seki-etal-2018-purely}, with most research focusing on attention-based encoder-decoders (AEDs). 
For meeting transcription, a well-studied framework is \textit{serialized output training} (SOT), wherein multiple references are serialized into a single prediction sequence based on special tokens~\cite{Kanda2020SerializedOT}. 
A detailed review of related work is presented in Section~\ref{sec:related}.

For the case of streaming multi-talker ASR, two methods have been proposed to extend neural transducers~\cite{Graves2012SequenceTW} for the task.
These methods are beneficial since neural transducers (using RNNs or transformers) have become the standard modeling technique for on-device speech recognition~\cite{he2019streaming, Wu2020StreamingTA, Li2019RNNT}.
The first method, exemplified by token-level SOT (t-SOT)~\cite{Kanda2022StreamingMA}, uses a conventional single-branch model similar to single-talker ASR, whereas the second method relies on a two-branch modeling strategy, as exemplified by \textit{streaming unmixing and recognition transducer} (SURT)~\cite{Lu2020StreamingEM} and \textit{multi-turn RNNT} (MT-RNNT)~\cite{Sklyar2021StreamingMA}.
In this paper, we focus on the latter class of models, and refer to them as SURT, but the same ideas should also be applicable to MT-RNNT.
The SURT model separates overlapping speech into multiple simultaneous \textit{branches} (or channels), each of which is transcribed by a shared transducer. 
It has been extended to handle long-form multi-turn recordings~\cite{Raj2021ContinuousSM,Sklyar2021MultiTurnRF}, and to jointly perform speaker identification~\cite{Lu2021StreamingMS}, endpointing~\cite{Lu2022EndpointDF}, and segmentation~\cite{Sklyar2022SeparatorTransducerSegmenterSR}. 

Although SURT is a promising framework for end-to-end multi-talker ASR, it suffers from several limitations. 
It was shown in~\cite{Raj2021ContinuousSM} that SURT performance often degrades on multi-turn sessions due to \textit{omission} and \textit{leakage} related errors. 
Here, omission refers to the case when an utterance is missed by all output branches, whereas leakage happens when a non-overlapping segment is transcribed on multiple branches.
Additionally, SURT requires training on long sessions with the transducer loss, which may be computationally prohibitive, or even infeasible using academic resources. 
Consequently, there exist no open-source recipes for training SURT-like models. 
Furthermore, all evaluations of SURT so far are limited to the setting of simulated meetings (such as LibriCSS), and it is not clear whether the models transfer well to real-world settings.

In this work, our objective is three-fold:
\begin{enumerate}[label=(\roman*),wide,labelwidth=!,labelindent=0pt]
    \item to improve the training efficiency of SURT models so that they are feasible to train on an academic budget;
    \item to analyze errors (such as those caused by omission and leakage), and develop methods to improve multi-talker ASR performance of SURT models; and
    \item to demonstrate the effectiveness of the resulting model on real meeting benchmarks.
\end{enumerate}
 
Towards these objectives, we propose systematic modifications of several aspects of SURT --- relating to the model design, network architecture, training mixture simulation, loss functions, and training schemes. 
We conduct ablation studies to demonstrate the impact of each of these design choices. 
For (iii), we demonstrate the efficacy of the resulting SURT 2.0 on three public benchmarks: LibriCSS, AMI, and ICSI. 
Across all settings, our model outperforms the original SURT and MT-RNNT in terms of WER performance, while being smaller and more training efficient. 
We release all training recipes and model checkpoints through the open-source \texttt{icefall} toolkit: \texttt{\url{https://desh2608.github.io/pages/surt2}}.
%

\section{Related Work}
\label{sec:related}

\subsection{Joint optimization for multi-talker ASR}
Early work on joint separation and ASR involved hybrid HMM-DNN models as the ASR backbone~\cite{Qian2017SingleChannelMS,Chang2018MonauralMS}. 
These models were often trained with auxiliary speaker information~\cite{Chang2018AdaptivePI} or using transfer learning from single-speaker acoustic models~\cite{Tan2018KnowledgeTI}. 
%
%
With the success and flexibility of end-to-end ASR systems~\cite{Graves2006ConnectionistTC, Graves2012SequenceTW,Lu2016OnTT, Chorowski2015AttentionBasedMF, Chiu2018StateoftheArtSR, he2019streaming, Li2019RNNT, li2021recent}, researchers quickly adapted these into jointly optimized multi-talker ASR pipelines~\cite{seki-etal-2018-purely, Chang2020EndToEndMS}. 
Similar training schemes --- speaker embeddings, curriculum learning, or knowledge distillation --- were used to improve these pipelines~\cite{Denisov2019EndtoEndMS,Zhang2020ImprovingES,Lin2022SeparatetoRecognizeJM}.
Multi-channel extensions of these models have also been proposed that seek to improve separation capabilities through neural beamforming techniques~\cite{Chang2019EndtoendMM,Shi2022TrainFS}. 
For the single-channel case, time-domain modeling has been used to improve separation and consequently benefit downstream ASR performance~\cite{vonNeumann2019EndtoEndTO}. 
However, most of these models were studied in the context of AEDs that do not perform streaming transcription\footnote{There are concurrent efforts to perform streaming ASR with AEDs using monotonic attention mechanisms~\cite{Inaguma2020EnhancingMM,Li2022TransformerBasedSA,Tsunoo2019TransformerAW}, but with limited adoption.}, and evaluated on the simple setting of fully-overlapping two-speaker mixtures (such as the WSJ-Mix dataset~\cite{Isik2016SingleChannelMS}). 
On the other hand, real-world multi-talker settings usually contain an arbitrary number of speakers and sparse overlaps.

\subsection{Serialized output training (SOT)}

SOT was first proposed as a method to use existing AED architectures for multi-talker ASR~\cite{Kanda2020SerializedOT}. 
It was extended to perform joint speaker counting and speaker identification using an auxiliary speaker inventory~\cite{Kanda2020JointSC,Kanda2020InvestigationOE}. 
A token-level variant of SOT, in conjunction with neural transducers, has shown strong performance on streaming multi-talker ASR~\cite{Kanda2022StreamingMA,Kanda2022StreamingSA}, and has also been combined with multi-channel front-ends~\cite{Kanda2022VarArrayMT} and large-scale pre-training~\cite{Kanda2021LargeScalePO}. 
An advantage of t-SOT is that it allows the same model and training scheme to be used for both single and multi-talker settings. 
The recently concluded M2MeT challenge used SOT as the baseline system~\cite{Yu2021M2MetTI,Yu2022SummaryOT}.

\subsection{Continuous speech separation}

Continuous speech separation (CSS) refers to the task of generating overlap-free speech signals from a continuous audio stream consisting of multiple utterances spoken by different people~\cite{Chen2020ContinuousSS}. 
This task has its origins in early work on \textit{unmixing transducers}~\cite{Wu2020AnEA,Yoshioka2018RecognizingOS}. 
CSS research has seen increased activity in the last few years due to interest in sparsely-overlapped meeting separation. 
The original model (which used PIT-based supervised training of BLSTM encoders) has been improved by leveraging better architectures such as Conformer~\cite{Chen2021ContinuousSS}, two-stage training~\cite{Wu2021InvestigationOP}, and large-scale semi-supervised and self-supervised learning~\cite{Wang2022LeveragingRC,Chen2022SpeechSW}.
While the original CSS used PIT-based training, other training methods such as recurrent selective attention network (RSAN)~\cite{Kinoshita2018ListeningTE,Zhang2021ContinuousSS} and Graph-PIT~\cite{vonNeumann2021GraphPITGP,vonNeumann2023SegmentLessCS} have also been investigated. 
Multi-channel extensions of CSS have been proposed using complex spatial features~\cite{Wang2020MultimicrophoneCS}, low-latency beamforming~\cite{Yoshioka2019LowlatencySC}, and direction-of-arrival (DOA) based source localization~\cite{Wang2021LocalizationBS}. 
The SURT model may be regarded as a jointly optimized version of CSS with transducer-based ASR.

\subsection{Other cascaded systems}

Recent editions of the CHiME challenge~\cite{barker2018fifth,Watanabe2020CHiME6CT} saw the use of cascaded multi-talker ASR systems that leverage guided source separation (GSS)~\cite{Boeddeker2018FrontendPF}. 
GSS is an unsupervised target-speaker extraction (TSE) method that relies on pre-computed speaker activities and blind source separation to perform front-end enhancement of overlapped speech signals. 
Since its proposal, block-online~\cite{Horiguchi2020BlockOnlineGS} and GPU-accelerated~\cite{Raj2022GPUacceleratedGS} variants have been proposed to speed up the method. 
GSS has been combined with strong back-end ASR to obtain state-of-the-art performance on the CHiME-5 dataset~\cite{Kanda2019GuidedSS}; however, the separation performance depends heavily on estimated speaker activity, and ASR results may degrade heavily if the segmentation is poor~\cite{Arora2020TheJM,Medennikov2020TheSS}. 
Recently, a spatial mixture model based meeting separation method was proposed which generalizes GSS and avoids some of its pitfalls~\cite{Boeddeker2022AnIS}.
Nevertheless, this approach is naturally offline and requires multiple microphone channels to perform well.
%
%
Other cascades of separation and ASR for multi-talker transcription have explored methods such as neural TSE~\cite{molkov2021AuxiliaryLF}, mixture-invariant training~\cite{Sivaraman2021AdaptingSS}, sequential neural beamforming~\cite{Raj2020IntegrationOS}, and cross-channel attention~\cite{Yu2022MFCCAMultiFrameCA}.  

\begin{figure*}
    \centering
    \includegraphics[width=0.8\linewidth]{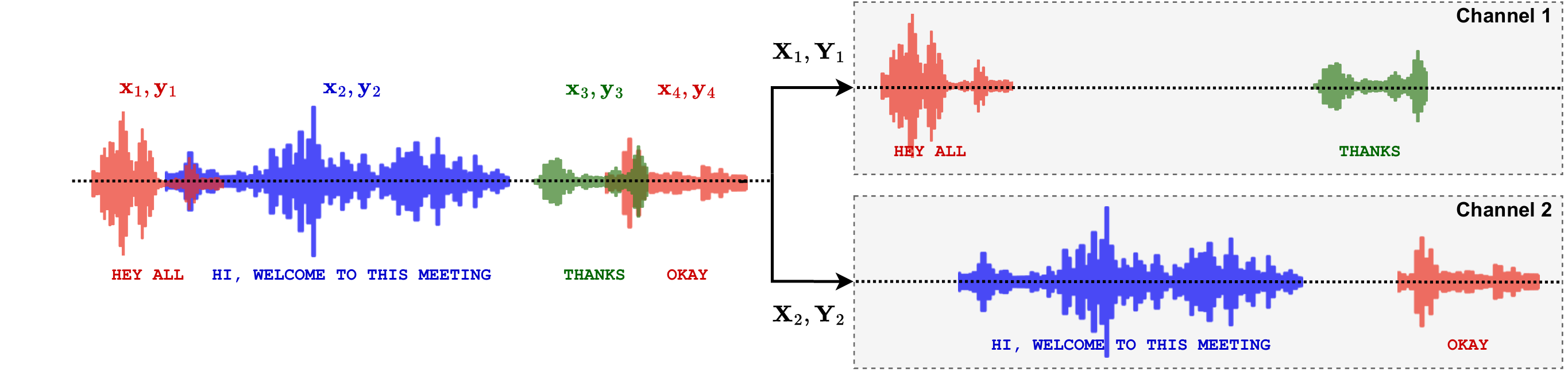}
    \caption{Example of heuristic error assignment training (HEAT). Each color represent a different speaker. The utterances ($\mathbf{x}_u$,$\mathbf{y}_u$) are assigned, in order of start times, to the next available channel. Such an assignment avoids the exponential complexity associated with permutation invariant training (PIT).}
    \label{fig:heat}
    \vspace{-1em}
\end{figure*}

\subsection{Multi-talker datasets}

Research on multi-talker ASR has benefited heavily from the availability of public corpora. 
These include early benchmarks such as AMI~\cite{Carletta2005TheAM}, ICSI~\cite{Janin2003TheIM}, and Sheffield Wargames~\cite{Fox2013TheSW}, and recently released datasets such as CHiME-5~\cite{barker2018fifth}, DiPCo~\cite{Segbroeck2019DiPCoD}, LibriCSS~\cite{Chen2020ContinuousSS}, AISHELL-4~\cite{Fu2021AISHELL4AO}, AliMeeting~\cite{Yu2021M2MetTI}, and RAMC~\cite{Yang2022OpenSM}. 
In this work, we evaluate our models on the LibriCSS, AMI, and ICSI meeting data.

\section{Preliminary}
\label{sec:review}

\subsection{Speech recognition with neural transducers}
\label{sec:asr}

In conventional single-talker ASR, audio features for a segmented utterance $\mathbf{X} \in \mathbb{R}^{T\times F}$ (where $T$ and $F$ denote the number of time frames and the input feature dimension, respectively) are provided as input to the system, and we are required to predict the transcript $\mathbf{y} = (y_1,\ldots,y_U)$, where $y_u \in \mathcal{V}$ denotes output units such as graphemes or word-pieces, and $U$ is the length of the label sequence. 
For the case of discriminative training, this requires computing the conditional likelihood $P(\mathbf{y}|\mathbf{X})$ (or its log for numerical stability). For inference, we search for $\hat{\mathbf{y}} = \text{arg}\max_{\mathbf{y}}P(\mathbf{y}|\mathbf{X})$, often in a constrained search space using greedy or beam search.
Transducers achieve this by marginalizing over the set of all alignments $\mathbf{a} \in \bar{\mathcal{V}}^{T+U}$, where $\bar{\mathcal{V}} = \mathcal{V}\cup \{\phi\}$ and $\phi$ is called the blank label.
Formally,
\begin{equation}
P(\mathbf{y}|\mathbf{X}) = \sum_{\mathbf{a}\in \mathcal{B}^{-1}(\mathbf{y})} P(\mathbf{a}|\mathbf{X}),
\label{eq:rnnt}
\end{equation}
where $\mathcal{B}$ is a deterministic mapping from an alignment $\mathbf{a}$ to an output sequence $\mathbf{y}$. 
In the original transducer, all alignments $\mathbf{a}$ consist of $\mathbf{y}$ interspersed with $T$ blank tokens, usually represented as a $T\times U$ lattice with $\phi$ on horizontal arcs and $\mathbf{y}$ on vertical arcs. 
Since there may be ${T+U \choose U}$ such paths on the lattice, some transducer variants (such as recurrent neural aligner~\cite{Sak2017RecurrentNA} or monotonic RNN-T~\cite{Tripathi2019MonotonicRN,Moritz2022AnIO}) restrict the number of non-blank tokens emitted per time step. 
External or internal alignments may also be used to further prune the lattice for marginalization~\cite{Mahadeokar2020AlignmentRS,Kuang2022PrunedRF}.  

\begin{figure*}[t]
\centering
\subfloat[\textbf{Original SURT}\label{fig:surt_old}]{%
       \includegraphics[trim={0.7cm 0cm 0 0},clip,width=0.49\linewidth]{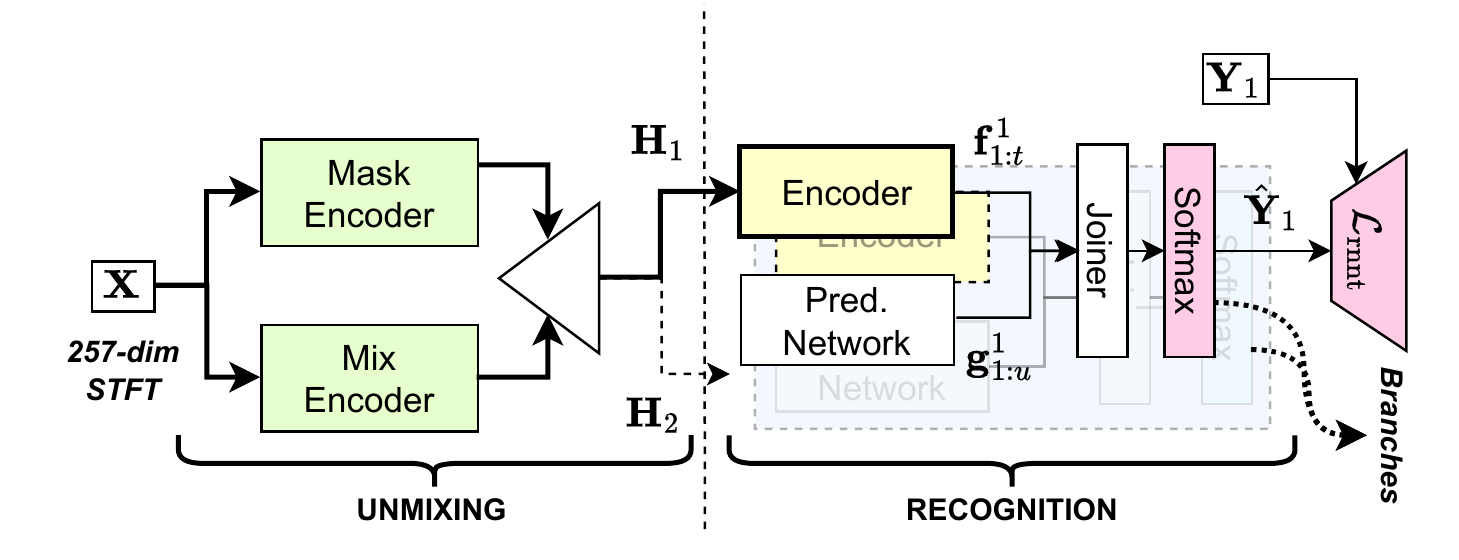}}
    \hfill
\subfloat[\textbf{SURT 2.0}\label{fig:surt_new}]{%
       \includegraphics[trim={0.75cm 0cm 0 0},clip,width=0.49\linewidth]{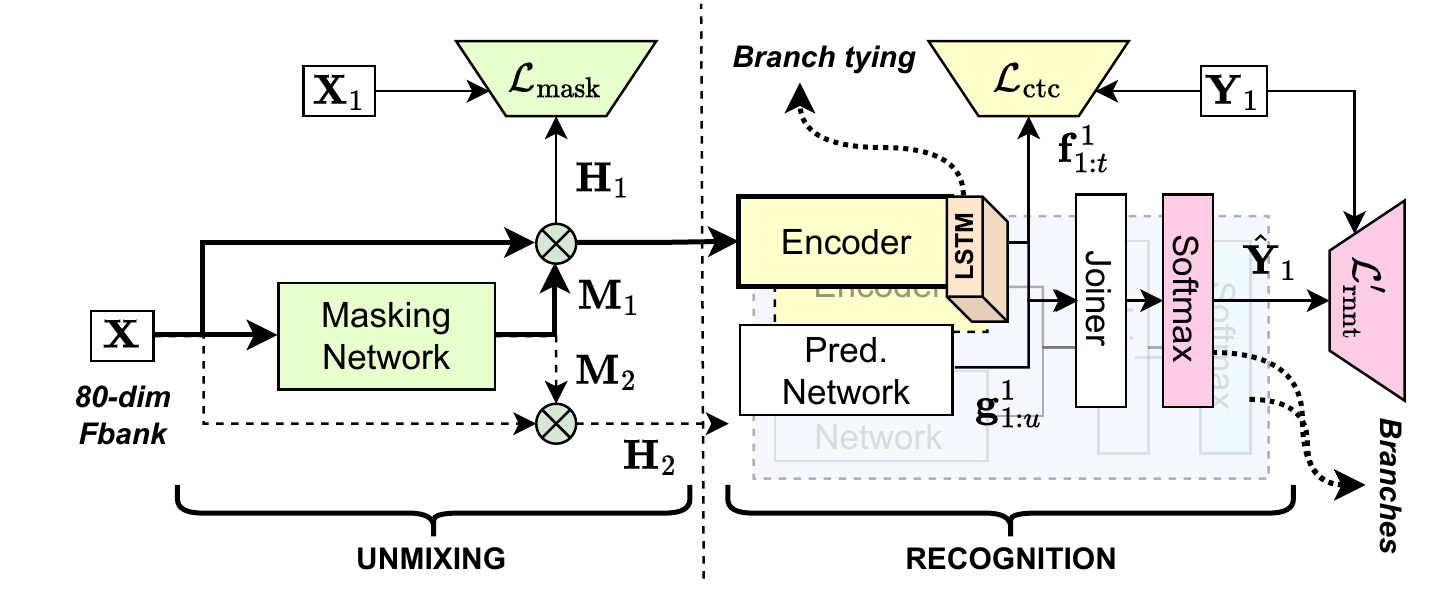}}
\caption{Overview of (a) original SURT, and (b) SURT 2.0. $\mathbf{X}$ denotes the input mixture. $\mathbf{X}_1$ and $\mathbf{Y}_1$ are concatenated sources and references for the first branch (only required during training). For SURT 2.0, the ``recognition'' component is pre-trained on single-speaker data. Table~\ref{tab:surt2} provides a summary of differences between the models.}
\label{fig:surt}
\vspace{-1em}
\end{figure*}

Transducers parameterize $P(\mathbf{a}|\mathbf{X})$ with an encoder, a prediction network, and a joiner (see ``recognition'' component in Fig.~\ref{fig:surt_old}).
The encoder maps $\mathbf{x}$ into hidden representations $\mathbf{f}_1^T$, while the prediction network maps $\mathbf{y}$ into $\mathbf{g}_1^U$.
The joiner combines the outputs from the encoder and the prediction network to compute logits $\mathbf{z}_{t,u}$ which are fed to a softmax function to produce a posterior distribution over $\bar{\mathcal{V}}$. 
Under assumptions of full context encoder, we can expand (\ref{eq:rnnt}) as
\begin{align}
    P(\mathbf{y}|\mathbf{X}) &= \sum_{\mathbf{a}\in \mathcal{B}^{-1}(\mathbf{y})} \prod_{t=1}^{T+U} P(\mathbf{a}_t|\mathbf{f}_1^T,\mathbf{g}_1^{u(t)-1}) \\
    &= \sum_{\mathbf{a}\in \mathcal{B}^{-1}(\mathbf{y})} \prod_{t=1}^{T+U} \mathrm{Softmax}(\mathbf{z}_{t,u(t)}),
\end{align}
where $u(t)\in\{1,\ldots,U\}$ denotes the index in the label sequence at time $t$. We will denote the log of this expression as $\mathcal{L}_{\text{rnnt}}(\mathbf{y},\mathbf{z})$ (or simply $\mathcal{L}_{\text{rnnt}}$) for the remainder of this paper.

\subsection{Multi-talker ASR with SURT}
\label{sec:mt-asr}

In multi-talker ASR, the input $\mathbf{X}\in\mathbb{R}^{T\times F}$ is an unsegmented mixture containing $N$ utterances from $K$ speakers, i.e., $\mathbf{X} = \sum_{n=1}^N \mathbf{x}_n$, where $\mathbf{x}_n$ is the $n$-th utterance ordered by start time, shifted and zero-padded to the length of $\mathbf{X}$. 
The desired output is $\mathbf{Y} = \{\mathbf{y}_n: 1\leq n \leq N\}$, where $\mathbf{y}_n$ is the reference corresponding to $\mathbf{x}_n$. 
If $N$ is small and fixed (e.g., $N=2$), a permutation-invariant training strategy may be used for this task. 
However, for general multi-talker scenarios, $N$ may be arbitrarily large.

Assuming at most two-speaker overlap, the \textit{heuristic error assignment training} (HEAT) paradigm~\cite{Lu2020StreamingEM} is used to create channel-wise references $\mathbf{Y}_1$ and $\mathbf{Y}_2$ by assigning $\mathbf{y}_n$'s to the first available channel, in order of start time (Fig.~\ref{fig:heat}). 
Formally, if $\zeta:n\rightarrow\{1,2\}$ maps utterances to channels, $\theta_n^{\mathrm{st}}$ and $\theta_n^{\mathrm{en}}$ denote the start and end times for utterance $n$, and $\theta_n^{\mathrm{st}}$ is monotonically increasing $\forall n$, we have
\begin{equation}
\label{eq:heat_labels}
    \zeta(n) = 
    \begin{cases}
    1, &\text{if}~~\theta_n^{\mathrm{st}} \geq \max_{i\in \zeta^{-1}(1)}\theta_i^{\mathrm{en}} \\
    2, &\text{otherwise},
    \end{cases}
\end{equation}
and $\zeta(n)$'s are assigned sequentially.
SURT estimates $\hat{\mathbf{Y}} = [\hat{\mathbf{Y}}_1,\hat{\mathbf{Y}}_2] = f_{\text{surt}}(\mathbf{X})$ as follows. 
First, an unmixing module computes $\mathbf{H}_1$ and $\mathbf{H}_2$ as
\begin{align}
\label{eq:surt}
& \mathbf{H}_1 = \mathbf{M} \ast \bar{\mathbf{X}}, \quad \mathbf{H}_2 = (\mathbbm{1} - \mathbf{M}) \ast \bar{\mathbf{X}}, \\
& \mathbf{M} = \sigma(\mathrm{MaskEnc}(\mathbf{X})) ~ \text{and} ~ \bar{\mathbf{X}} = \mathrm{MixEnc}(\mathbf{X}), \nonumber
\end{align}
where $\bar{\mathbf{X}},\mathbf{M},\mathbbm{1}\in \mathbb{R}^{T\times D}$ (for latent dim. $D$) is a matrix of ones, $\sigma$ is the sigmoid function, and $\ast$ is Hadamard product. 
$\mathbf{H}_1$ and $\mathbf{H}_2$ are fed into a transducer-based ASR, producing logits $\mathbf{Z}_1$ and $\mathbf{Z}_2$. 
Finally,
\begin{equation}
\label{eq:heat}
\mathcal{L}_{\text{heat}} = \mathcal{L}_{\text{rnnt}}(\mathbf{Y}_1, \mathbf{Z}_1) + \mathcal{L}_{\text{rnnt}}(\mathbf{Y}_2, \mathbf{Z}_2),
\end{equation}
where $\mathcal{L}_{\text{rnnt}}$ is the standard RNN-T loss~\cite{Graves2012SequenceTW}. 
SURT performs \textit{speaker-agnostic} transcription, and is evaluated using an optimal reference combination WER (ORC-WER)~\cite{Sklyar2021MultiTurnRF}, as described in Section~\ref{sec:evaluation}.

\section{Methodology: SURT 2.0}
\label{sec:method}

Multi-talker ASR with SURT requires the model to perform well on three challenging sub-tasks:
\begin{itemize}[itemindent=0em]
\item[\encircle{1}]continuous separation of sparsely overlapped speech; \item[\encircle{2}]long-form speech recognition; and
\item[\encircle{3}]modeling quick turn-taking among multiple speakers. 
\end{itemize}

The failure cases in SURT can be linked to model degeneration on one or more of these sub-tasks. 
For instance, \textit{omission} and \textit{leakage} errors may be attributed to \encircle{1}, while high error rates on quick turn-taking scenarios with short silences (see Section~\ref{sec:results}) may be caused by \encircle{3}. 
Deletion errors on long sequences may be caused by both \encircle{1} and \encircle{2}. 
SURT~2.0 contains several modifications aimed at addressing each of these sub-tasks, and to improve training efficiency of the model. 
Our modifications are summarized in Table~\ref{tab:surt2}, and the resulting model is shown in Fig.~\ref{fig:surt_new}.

\begin{table}[t]
\centering
\caption{Overview of differences between original SURT and SURT 2.0 (for training on LibriSpeech mixtures).}
\label{tab:surt2}
\adjustbox{max width=\linewidth}{
\begin{tabular}{@{}lll@{}}
\toprule
 & \textbf{Original SURT} & \textbf{SURT 2.0} \\
\midrule
\textit{Modeling} & \begin{tabular}[c]{@{}l@{}}Features: 257-dim STFT\\Mask/mix encoders for unmixing\end{tabular} & \begin{tabular}[c]{@{}l@{}}Features: 80-dim fbank\\ Masking network for unmixing\end{tabular} \\
\midrule

\textit{Architecture} & \begin{tabular}[c]{@{}l@{}}Mask/mix encoder: Conv2D\\Encoder: DP-LSTM / DP-Transformer \\ Predictor: LSTM\end{tabular} & 
\begin{tabular}[c]{@{}l@{}}Mask encoder: DP-LSTM\\Encoder: Branch-tied zipformer \\ Predictor: Conv1D\end{tabular}\\
\midrule

\textit{Loss function} & $\mathcal{L}_{\mathrm{rnnt}}$ & $\mathcal{L}^{\prime}_{\mathrm{rnnt}} + \lambda_{\mathrm{ctc}}\mathcal{L}_{\mathrm{ctc}} + \lambda_{\mathrm{mask}}\mathcal{L}_{\mathrm{mask}}$ \\
\midrule

\begin{tabular}[l]{@{}l@{}}\textit{Mixture}\\\textit{simulation}\end{tabular} & \begin{tabular}[c]{@{}l@{}}Source: LibriSpeech utterances \\2 speakers, $\leqslant$ 4 turns\\Avg. duration (s): 25.5$\pm$5.0$^{\dagger}$ \\ Far-field: simulated RIRs \end{tabular} &
\begin{tabular}[c]{@{}l@{}}Source: LibriSpeech segments \\2--3 speakers, $\leqslant$ 9 turns \\Avg. duration (s): 16.2$\pm$4.7 \\ Far-field: real RIRs \end{tabular}\\
\midrule

\begin{tabular}[c]{@{}l@{}}\textit{Pre-training} \\ \textit{\& adaptation}\end{tabular} & None & \begin{tabular}[c]{@{}l@{}}Transducer pre-training \\ Adaptation using in-domain data\end{tabular} \\
\bottomrule
\multicolumn{3}{@{}l}{$^\dagger$Based on our estimate using the simulation configuration.} \\
\end{tabular}
}
\end{table}

\subsection{Modeling}
\label{sec:model}


For ``unmixing'' (or separation), SURT used dual mix/mask encoders that projected input 257-dim STFTs into high dimensional representations, as shown in (\ref{eq:surt}). 
Clearly, this design constrains SURT to have exactly two output branches, and the separated features are not interpretable. 
Instead, we use 80-dimensional log Mel filter-banks as inputs, and replace the mix/mask encoders with a simple masking network that can generate arbitrary number of masks. 
Formally, given output channel count $C$, our unmixing module generates masks
\begin{equation}
[\mathbf{M}_1,\ldots,\mathbf{M}_C]^T = \mathrm{MaskNet}(\mathbf{X}),
\end{equation}
where $\mathbf{M}_c \in \mathbb{R}^{T\times F}$. 
These masks are applied to the input $\mathbf{X}$ to obtain channel-specific features: $\mathbf{H}_c = \mathbf{M}_c \ast \mathbf{X}$. 
Such a design has three advantages: (i) it allows the use of arbitrary number of output branches $C$, (ii) the masked representations $\mathbf{H}_c$ are interpretable as clean features, and (iii) it allows pre-training of the recognition module on single-speaker speech. 
We will describe (ii) and (iii) further in subsequent sections. 

Our recognition module is similar to SURT, other than architectural changes (Section~\ref{sec:arch}). 
Channel-wise features $\mathbf{H}_c$ are fed into an encoder which generates hidden representations $\mathbf{f}_{1:T}^c$. 
The corresponding label sequence $\mathbf{Y}_c$ is fed into a prediction network as per the HEAT training strategy, generating hidden representations $\mathbf{g}_{1:U}^c$. 
A joiner combines $\mathbf{f}_{1:T}^c$ and $\mathbf{g}_{1:U}^c$ to generate logits $\mathbf{z}_{t,u}^c$. 
The parameters of the encoder, prediction network, and joiner are shared among all the output branches, as shown in Fig.~\ref{fig:surt}. 
Further details about model hyperparameters are given in Section~\ref{sec:implementation}.

\subsection{Network architecture}
\label{sec:arch}


\begin{figure}[t]
    \centering
    \includegraphics[width=0.9\linewidth]{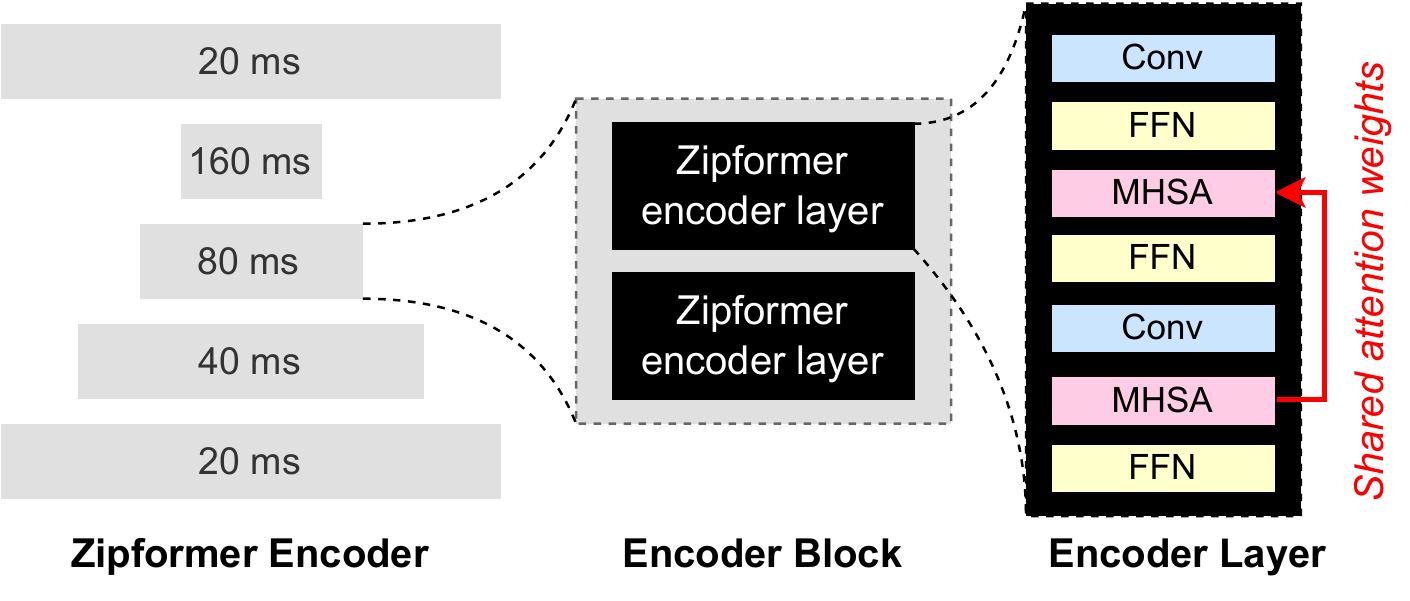}
    \vspace{-1em}
    \caption{Illustration of the zipformer encoder architecture. The encoder contains multiple ``blocks'' running at different frame rates (left). Each block contains several ``layers'' (middle), and each layer performs self-attention twice with shared attention weights (right).}
    \label{fig:zipformer}
\end{figure}

We made several changes to the network architecture of SURT modules. 
For the masking network, we use dual-path LSTMs (DP-LSTMs) instead of 2-D convolutions in order to improve long range modeling capability for unmixing~\cite{Luo2019DualPathRE}. 
Second, in the encoder, we replace the DP-LSTMs (or DP-Transformers) with the recently proposed zipformer~\cite{zipformer}. 
As shown in Fig.~\ref{fig:zipformer}, the zipformer encoder contains multiple encoder blocks running at different rates, with the middle ones more strongly down-sampled (by up to a factor of 8). 
This makes training more efficient as there are fewer frames to evaluate. 
Each block may contain one or more encoder ``layers'' operating at the same frame rate.
Each layer performs self-attention twice with shared attention weights, and a trainable bypass is introduced for each layer dimension.
We also propose ``branch tying'' of the encoder outputs to \textit{jointly} learn representation across all branches, i.e.,
\begin{equation}
\label{eq:bt}
    [\hat{\mathbf{h}}_1^{\text{enc}},\ldots,\hat{\mathbf{h}}_C^{\text{enc}}]^T = \mathrm{LSTM}\left( [\mathbf{h}_1^{\text{enc}},\ldots,\mathbf{h}_C^{\text{enc}}]^T \right).
\end{equation}

The motivation for branch-tied encoders is to reduce errors from \textit{omission} and \textit{leakage}, which usually happen when output branches do not communicate. 
Finally, we use a ``stateless'' prediction network~\cite{Ghodsi2020RnnTransducerWS} instead of LSTM layers in the original SURT. 
In addition to improving computational efficiency, we conjecture that a stateless network should also be better suited to handle quick turn-taking (sub-task~\encircle{3} described earlier).
We set the segment length for the bi-directional intra-LSTM (of the DP-LSTM network) equal to the chunk size of the causal Zipformer encoder; this is the overall latency of the SURT model.

\subsection{Training objective}
\label{sec:loss}


SURT was trained end-to-end with the transducer loss~\cite{Graves2012SequenceTW}, with the utterance-wise permutation resolved by HEAT (Fig.~\ref{fig:heat}).
However, the transducer loss suffers from high memory usage, since it requires marginalization over a logit tensor of size ($B,T,U,D$). 
Even with an efficient implementation~\cite{Li2019RNNT}, this severely limits the training sequence length --- for example, in~\cite{Raj2021ContinuousSM}, the authors trained SURT with mixtures containing at most 4 speaker turns. 
To remedy this issue, we replace the full-sum transducer loss with the recently proposed \textit{pruned} transducer loss, which prunes the alignment lattice using a simple linear joiner before computing the full sum on the pruned lattice~\cite{Kuang2022PrunedRF}. 
%

Recall from Section~\ref{sec:asr} that the encoder and the prediction network generate hidden representations $\mathbf{f}_1^T$ and $\mathbf{g}_1^U$, respectively.
In pruned transducer, these are first projected to $\mathbb{R}^{|\bar{\mathcal{V}}|}$, and denoted as $\hat{\mathbf{f}}_1^T$ and $\hat{\mathbf{g}}_1^U$, respectively. 
A simple additive joiner is then used to compute \textit{trivial} logits $\hat{\mathbf{z}}_{t,u}$ as
\begin{align}
   & \hat{\mathbf{z}}_{t,u} = \hat{\mathbf{f}}_{t} + \hat{\mathbf{g}}_{u} - \hat{\mathbf{z}}^{\mathrm{norm}}_{t,u}, \\
    \text{where}~~ &\hat{\mathbf{z}}^{\mathrm{norm}}_{t,u} = \log \sum_{v} \exp \left( \hat{\mathbf{f}}_{t} + \hat{\mathbf{g}}_{u} \right). \label{eq:log_matrix}
\end{align}
Here, equation~(\ref{eq:log_matrix}) can be interpreted as log-space matrix multiplication, and is easy to implement through simple matmul operations. 
Gradients from this simple joiner are used to compute locally optimal pruning bounds for the lattice\footnote{The pruning is ``locally'' optimal in the sense that the optimization is performed per-frame, as opposed to a ``globally'' optimal treatment which would consider the whole path through the lattice.
This local strategy may result in path discontinuity that is later resolved through adjustments.}, and the full joiner output, $\mathbf{z}_{t,u}$, is only computed on the pruned lattice. 
We used the open-source implementation available in \texttt{k2}: \texttt{\url{https://github.com/k2-fsa/k2}}.

Additionally, we use auxiliary loss functions to regularize SURT training and improve separation of sparsely overlapped speech. 
For the former, we add a connectionist temporal classification (CTC) loss~\cite{Graves2006ConnectionistTC}, which has been shown to provide regularization capabilities due to monotonicity in alignments~\cite{Kim2016JointCB,Sudo20224DAJ}. This is given as
\begin{equation}
\label{eq:ctc}
\mathcal{L}_{\text{ctc}} = \log \sum_{\mathbf{a}\in \mathcal{B}_{\text{ctc}}^{-1}(\mathbf{y})} \prod_t P(\mathbf{a}_t|\mathbf{f}_1^T),
\end{equation}
where $\mathbf{a}$ is a $T$-length sequence that deterministically maps to $\mathbf{y}$ through transformation $\mathcal{B}_{\text{ctc}}$, which removes repeated tokens and $\phi$.
To improve mask estimation, we apply a masking loss directly on the outputs $\mathbf{H}_c$ generated by the mask encoder as
\begin{equation}
\label{eq:mask}
    \mathcal{L}_{\text{mask}} = \sum_{c \in C}\mathrm{MSE}(\mathbf{H}_c, \mathbf{X}_c),
\end{equation}
where $\mathrm{MSE}$ denotes mean-squared error, and $\mathbf{X}_c$ is obtained by summing clean inputs $\mathbf{x}_u$ assigned to branch $c$ (Fig.~\ref{fig:heat}). 
The overall training objective is given as
\begin{equation}
    \label{eq:loss}
    \mathcal{L} = \mathcal{L}^{\prime}_{\text{rnnt}} + \lambda_{\text{ctc}}\mathcal{L}_{\text{ctc}} + \lambda_{\text{mask}}\mathcal{L}_{\text{mask}},
\end{equation}
where $\mathcal{L}^{\prime}_{\text{rnnt}}$ denotes the pruned transducer loss and $\lambda$'s are hyperparameters.

\begin{algorithm}[t]
\DontPrintSemicolon
  
  \KwInput{$\mathcal{X}$, $\mathcal{M}$, $K$, $T$}
  \KwOutput{$\mathcal{S}$}
  
  $D_{=\text{spk}}, D_{\neq\text{spk}}, D_{\text{ovl}}, \mathcal{S}$ = $\phi$
  
  \tcp*[l]{Fit distributions to $\mathcal{M}$}
  \For{$M$ in $\mathcal{M}$}{
    \For{$i$ in range($|M|$)}{
      $t$ = $M_i.start - M_{i-1}.end$
      
      \eIf{$M_i$.spk == $M_{i-1}$.spk}{
        $D_{=\text{spk}}$ = $D_{=\text{spk}} \cup \{t\}$ 
      }{\eIf{$t > 0$}{
          $D_{\neq\text{spk}}$ = $D_{\neq\text{spk}} \cup \{t\}$ 
        }{
          $D_{\text{ovl}}$ = $D_{\text{ovl}} \cup \{-t\}$ 
        }
      }
    }
  }

  $P_{\text{ovl}}$ = $\frac{|D_{\text{ovl}}|}{|D_{\neq\text{spk}}|+|D_{\text{ovl}}|}$; $D_{\ast}$ = histogram($D_{\ast}$)

  \tcp*[l]{Generate mixtures using $\mathcal{X}$}
  $\mathcal{X}$ = $\{\mathcal{X}_1,\ldots,\mathcal{X}_S\}$ \tcp*[l]{speaker wise bucketing of $\mathcal{X}$}

  \While{any($|\mathcal{X}_s| > 0$)}{
    \tcp*[l]{Select speakers}
    $k \leftarrow \text{sample}(K)$; $\mathcal{X}_{s_1},\ldots,\mathcal{X}_{s_k} \leftarrow \text{sample}(\mathcal{X})$

    \tcp*[l]{Select utterances for each speaker}
    \For{$i$ in range($k$)}{
      $U_{s_k} \leftarrow \text{sample}(\mathcal{X}_{s_k}), \text{s.t.} \left(\sum_{u\in U_{s_k}} u.dur\right) < T$ 

      $\mathcal{X}_{s_k} \leftarrow \mathcal{X}_{s_k}\setminus U_{s_k}$
    }

    $U = \text{shuffle}(U_{s_1},\ldots,U_{s_k})$; offset = 0

    \tcp*[l]{Get offsets for each utterance}
    $\mathcal{S}_{\mathrm{cur}}\leftarrow \phi$ \tcp*[l]{initialize empty mixture}
    
    \For{$i$ in range($|U|$)}{
      \eIf{$U_i$.spk == $U_{i-1}$.spk}{
        ot = sample($D_{=\text{spk}}$)
      }{\eIf{Bernoulli($P_{\text{ovl}}$>0.5)}{
        ot = --sample($D_{\text{ovl}}$)
      }{
        ot = sample($D_{\neq\text{spk}}$)
      }
      }
      offset = offset + ot
      
      $\mathcal{S}_{\mathrm{cur}} = \mathcal{S}_{\mathrm{cur}} \cup \{U_i,\text{offset}\}$
    }
    $\mathcal{S} = \mathcal{S} \cup \mathcal{S}_{\mathrm{cur}}$
  }

\caption{Training mixture simulation}
\label{alg:mixture}
\end{algorithm}

\begin{figure}[t]
    \centering
    \includegraphics[width=\linewidth]{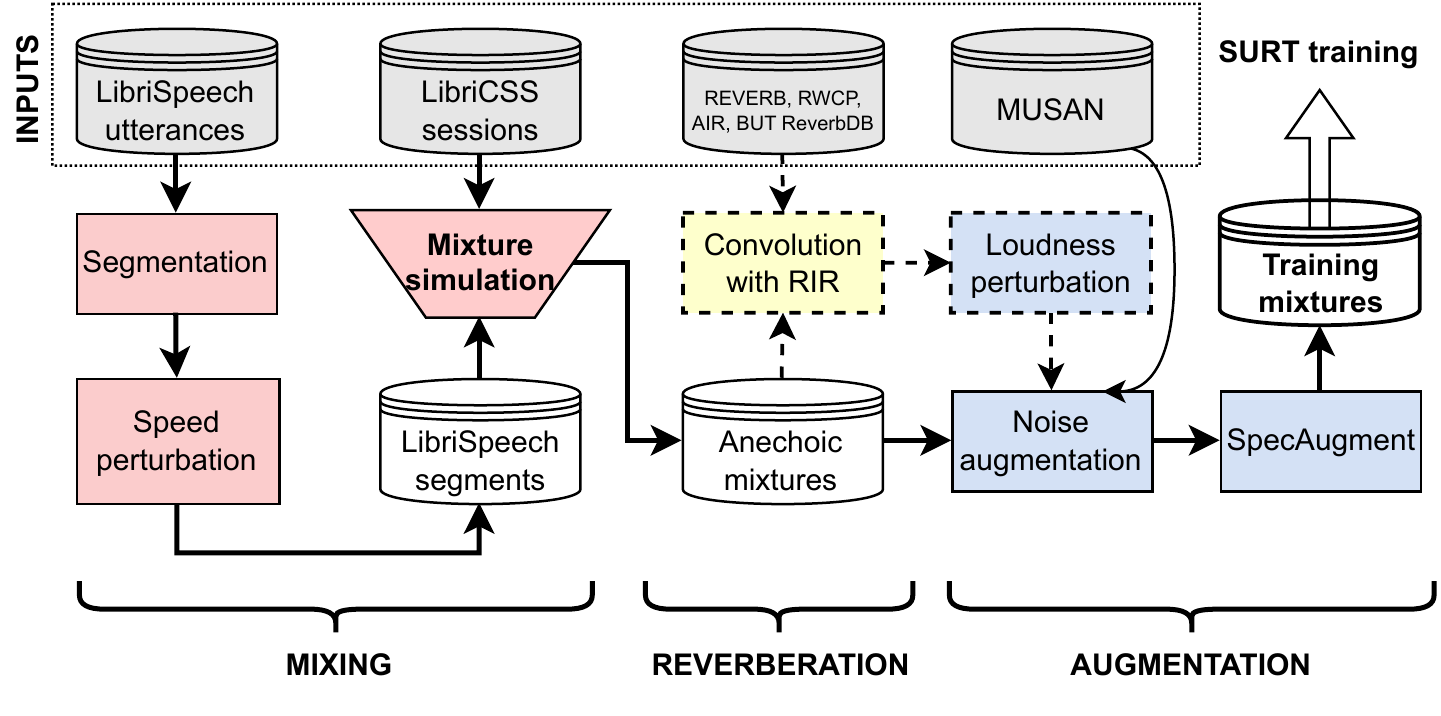}
    \vspace{-2em}
    \caption{Mixture simulation workflow for LibriSpeech-based training. Gray cylinders denote external data used during simulation: LibriSpeech utterances, statistics from LibriCSS, real RIRs, and MUSAN noises. The simulation workflow can be conceptually divided into three phases: (i) \textit{mixing}, which creates anechoic meetings, (ii) \textit{reverberation}, which convolves with RIRs, and (iii) \textit{augmentation}, which perturbs the reverberant mixtures with various schemes. The dotted path is optional for anechoic training.}
    \label{fig:simulation}
\end{figure}

\subsection{Mixture simulation}
\label{sec:sim}


The original SURT was trained on synthetic mixtures of \textit{full} utterances, which may result in prohibitively long sequences; for e.g., LibriSpeech \texttt{train} has an average duration of 12.4s. 
Furthermore, segment overlaps were arbitrarily determined to satisfy overlap ratio constraints, which may not be representative of the target sessions. 
%

For SURT 2.0, we use \textit{sub-segments} instead of full utterances as the source for mixture simulation, so that resulting mixtures are shorter while retaining multiple turns of conversation.
These sub-segments are obtained using word-level alignment information, by breaking up the utterances at pauses longer than a threshold $\tau$.
As an example, using $\tau=0.2$ for LibriSpeech resulted in sub-segments that were 2.8s on average. 
This allowed each training session to contain up to 9 turns of conversation while still being 36.5\% shorter than the original training mixtures (Table~\ref{tab:surt2}). 
Second, we learn histograms of pause/overlap distribution statistics from the target sessions, and sample from these distributions for mixing the segments.
Such a strategy has been successfully applied to improve end-to-end neural diarization~\cite{Landini2022FromSM}. 
Our mixture simulation algorithm is described in Algorithm~\ref{alg:mixture}, and is similar to the conversation simulation algorithm from~\cite{Landini2022FromSM}. 
We assume that the input to the algorithm is the source segments $\mathcal{X}$, target sessions $\mathcal{M}$ (to learn statistics), maximum number of speakers $K$ in each mixture, and maximum duration $T$ of a speaker in a mixture. 
The algorithm returns the training mixtures $\mathcal{S}$. 
For training, the input speech mixture $\mathbf{X}$ is obtained by digitally adding the utterances in $\mathcal{S}$ with the specified offsets.
The utterance-wise labels $\mathbf{y}_n$, along with $\theta_n^{\mathrm{st}}$ and $\theta_n^{\mathrm{en}}$ (uniquely determined by the offset and duration of utterance $n$) are used to obtain the reference transcripts for both channels using equation~(\ref{eq:heat_labels}).
We optionally convolve $\mathcal{S}$ with real room impulse responses (RIRs) to train models for far-field reverberant conditions.
The overall simulation workflow is shown in Fig.~\ref{fig:simulation}.


\subsection{Pre-training \& adaptation}
\label{sec:pretrain}


Using sub-segments for mixture simulation improves training efficiency while allowing multiple speaker turns; however, it creates a train-test mismatch for duration of individual segments, which could degrade the model's performance on sub-task \encircle{2}, i.e., to recognize long utterances from the same speaker. 
We solve this problem by pre-training the transducer module on single-speaker utterances (e.g., on LibriSpeech \texttt{train} set).
Such a pre-training strategy also decouples the tasks of learning to separate from learning to transcribe, and helps the SURT model converge faster. 
Recall from Section~\ref{sec:model} that this pre-training is possible in SURT 2.0 because our masking network generates masked filter-banks instead of high-dimensional latent representations.

Despite convolving with real RIRs, the acoustic characteristics of the mixtures used to train SURT may still be mismatched from real meeting recordings. As a final step of SURT training, we perform model adaptation by training on in-domain data for a small number of iterations. 

\section{Experimental Setup}
\label{sec:setup}

%

\subsection{Data}
\label{sec:data}

We performed evaluations on three publicly-available meeting datasets in English: LibriCSS, AMI, and ICSI. The summary statistics for these datasets are shown in Table~\ref{tab:stats}.

\textbf{LibriCSS} consists of multi-channel audio recordings of 8-speaker \textit{simulated} conversations that were created by combining utterances from the LibriSpeech \texttt{test-clean} set~\cite{Panayotov2015LibrispeechAA} and playing in real meeting rooms. 
It comprises 10 one-hour long sessions, each made up of six 10-minute ``mini sessions'' that have different overlap ratios (ranging from 0\% to 40\%). 
We show results on \textit{anechoic} and \textit{replayed} versions of LibriCSS, where the former consists of digitally mixed utterances, and the latter is the replayed sessions (we choose the first channel for evaluation). 
%
%
We used session 0 as the \texttt{dev} set and the remaining for \texttt{test}, following previous work~\cite{Raj2020IntegrationOS}.

\textbf{AMI} consists of 100 hours of recorded meetings containing 4 or 5 speakers per session~\cite{Carletta2005TheAM}.
Sessions were recorded on close-talk (headset and lapel) microphones, as well as 2 linear arrays each containing 8 mics.
We used three different mic settings for our experiments: IHM-Mix (digitally mixed individual headset mics), SDM (first channel of array-1), and MDM (beamformed array-1), where the last setting uses officially provided beamformed recordings~\cite{Mir2007AcousticBF}. 

\textbf{ICSI} data comprises around 72 hours of natural, meeting-style overlapped speech recorded at International Computer Science Institute (ICSI), Berkley.
Speech was captured using close-talk headsets and ad-hoc distant microphones.
The original data did not provide any partitions, so we use the speaker-disjoint partitions suggested in~\cite{Renals2014NeuralNF}.
We performed evaluations on the IHM-Mix and SDM settings --- for SDM, we selected the third distant mic, similar to the Kaldi recipe\footnote{\url{https://github.com/kaldi-asr/kaldi/tree/master/egs/icsi}}.

\begin{table}[t]
\centering
\caption{Statistics of datasets used for evaluations. The $k$-speaker durations are in terms of fraction of total speaking time.}
\label{tab:stats}
\adjustbox{max width=\linewidth}{
\begin{tabular}{@{}lrrrrrrrr@{}}
\toprule
\multirow{2}{*}{} & \multicolumn{2}{c}{\textbf{LibriCSS}} & \multicolumn{3}{c}{\textbf{AMI}} & \multicolumn{3}{c}{\textbf{ICSI}} \\
\cmidrule(r{5pt}){2-3} \cmidrule(l{4pt}){4-6} \cmidrule(l{4pt}){7-9}
 & \textbf{Dev} & \textbf{Test} & \textbf{Train} & \textbf{Dev} & \textbf{Test} & \textbf{Train} & \textbf{Dev} & \textbf{Test} \\ 
\midrule
\textbf{Duration (h:m)} & 1:00 & 9:05 & 79:23 & 9:40 & 9:03 & 66:38 & 2:16 & 2:45 \\
\textbf{Num. sessions} & 6 & 54 & 133 & 18 & 16 & 70 & 2 & 3 \\
\textbf{Silence (\%)} & 6.2 & 6.7 & 18.1 & 21.5 & 19.6 & 55.2 & 25.9 & 25.9 \\
\textbf{1-speaker (\%)} & 81.3 & 81.2 & 75.5 & 74.3 & 73.0 & 82.1 & 90.3 & 84.9 \\
\textbf{2-speaker (\%)} & 18.6 & 18.5 & 21.1 & 22.2 & 21.0 & 15.7 & 9.0 & 13.6 \\
\textbf{>2-speaker (\%)} & 0.1 & 0.4 & 3.4 & 3.5 & 6.0 & 2.2 & 0.7 & 1.4 \\
\bottomrule
\end{tabular}}
\end{table}

\subsection{Evaluation}
\label{sec:evaluation}

Fig.~\ref{fig:heat} shows that SURT generates speaker-agnostic transcription with no strict correspondence between speaker and channel. 
This makes it impossible to evaluate SURT with the conventional word error rate (WER) metric used for single-speaker ASR systems.
\cite{Raj2021ContinuousSM} and \cite{Sklyar2021MultiTurnRF} independently proposed similar metrics that computed the minimum word error rate (WER) based on an optimal assignment of references to the channels, and this was termed optimal reference combination WER (ORC-WER) in~\cite{Sklyar2021MultiTurnRF}.
In both studies, the authors remarked that computation of this metric was exponential in the number of reference utterances.
Von Neumann et al.~\cite{vonNeumann2022OnWE} proposed a polynomial-time implementation of ORC-WER using multi-dimensional Levenshtein distance, and released it through the \texttt{meeteval}\footnote{\url{https://github.com/fgnt/meeteval}} toolkit.
They showed that ORC-WER is a lower bound on the popular concatenated minimum-permutation WER (cpWER)~\cite{Watanabe2020CHiME6CT}, and becomes equal to the cpWER when no speaker errors are present.
We used their open-source implementation for evaluating our SURT models.
%
%
%
In the remainder of this paper, we will use the abbreviation WER to actually mean ORC-WER, unless explicitly mentioned.
For LibriCSS, we report WERs on the officially provided (approximately 1 min. long) ``segments,'' following prior work on continuous input evaluation~\cite{Chen2020ContinuousSS,Chen2021ContinuousSS,Raj2021ContinuousSM}. 
For AMI and ICSI, we use utterance-group based evaluation similar to \cite{Kanda2021LargeScalePO,Huang2022AdaptingSM}.
This results in \texttt{dev}/\texttt{test} segments of duration 7.0s/8.4s and 2.6s/2.9s for AMI and ICSI, respectively.

\subsection{Implementation details}
\label{sec:implementation}

\begin{table*}[t]
\centering
\caption{Comparison with original SURT and MT-RNNT models on ``anechoic'' and ``replayed'' versions of LibriCSS \texttt{test} set. The SURT 2.0 models were decoded using beam search with a beam size of 4.}
\label{tab:results}
\adjustbox{max width=\linewidth}{
\begin{tabular}{@{}lc@{\hskip 2em}ccccccr@{}@{\hskip 2em}@{}ccccccr@{}}
\toprule
\multirow{2}{*}{\textbf{Model}} & \multirow{2}{*}{\textbf{Size (M)}} & \multicolumn{7}{c}{\textbf{Anechoic}} & \multicolumn{7}{c}{\textbf{Replayed}} \\
\cmidrule(r{12pt}){3-9} \cmidrule(l{1pt}){10-16}
& & \textbf{0L} & \textbf{0S} & \textbf{OV10} & \textbf{OV20} & \textbf{OV30} & \textbf{OV40} & \textbf{Avg.} & \textbf{0L} & \textbf{0S} & \textbf{OV10} & \textbf{OV20} & \textbf{OV30} & \textbf{OV40} & \textbf{Avg.} \\ \midrule

Original SURT~\cite{Raj2021ContinuousSM} & 42.9 & 6.9 & 18.9 & 19.6 & 21.9 & 23.9 & 28.7 & 20.0 & 9.3 & 21.1 & 21.2 & 25.9 & 28.2 & 31.7 & 22.9\\
Multi-turn RNN-T~\cite{Sklyar2021MultiTurnRF} & 81.0 & -- & -- & -- & -- & -- & -- & -- & 14.8 & 14.5 & 18.0 & 25.8 & 30.3 & 32.3 & 22.6 \\
\midrule

SURT 2.0 (Base) & 26.7 & 5.2 & 4.7 & 14.2 & 17.8 & 21.3 & 23.0 & 14.4 & 6.7 & 8.2 & 23.0 & 27.1 & 28.5 & 31.8 & 20.9 \\
$\hookrightarrow$ w/ \texttt{dev} adaptation & 26.7 & 5.1 & 4.2 & 13.7 & 18.7 & 20.5 & \textbf{20.6} & 13.8 & 6.8 & 7.2 & 21.4 & 24.5 & 28.6 & 31.2 & 20.0 \\
\midrule

SURT 2.0 (Large) & 37.9 & \textbf{4.6} & \textbf{3.8} & 14.9 & 17.3 & 19.1 & 23.9 & 13.9 & \textbf{5.9} & 7.8 & 21.2 & 25.7 & 27.8 & 29.9 & 19.7 \\
$\hookrightarrow$ w/ \texttt{dev} adaptation & 37.9 & \textbf{4.6} & \textbf{3.8} & \textbf{12.7} & \textbf{14.3} & \textbf{16.7} & 21.2 & \textbf{12.2} & 6.4 & \textbf{6.9} & \textbf{17.9} & \textbf{19.7} & \textbf{25.2} & \textbf{25.5} & \textbf{16.9} \\
\bottomrule
\end{tabular}
}
\vspace{-1.5em}
\end{table*}

\noindent
\textbf{Network architecture.}
We experimented with two variants of SURT --- \textit{base} and \textit{large}. The \textit{base} model contains four 256-dim DP-LSTM layers trained with chunk width randomization~\cite{Raj2021ContinuousSM} as the masking network. 
The encoder consists of 5 zipformer blocks with 2 self-attention layers per block.
%
Each block consists of a 192-dim attention distributed across 8 heads, and a 768-dim feed-forward layer. 
Downsampling factors of (1,2,4,8,2) were used in the zipformer blocks. 
The prediction network contains a single 512-dim Conv1D layer. 
The \textit{large} model contains 6 layers in the masking network, and (2,4,3,2,4) self-attention layers in the 5 zipformer blocks.
The chunk size for the intra-LSTM and the Zipformer is set to 32 frames, resulting in a modeling latency of 320 ms.

\noindent
\textbf{Training data.}
For LibriCSS experiments, we first created anechoic mixtures, \textbf{LSMix-clean}, using each speed-perturbed LibriSpeech train sub-segment ($\tau=0.2$; cf. \S~\ref{sec:sim}) once, resulting in approx. 2200h of training data.
For this, we set $\mathcal{M}$ as the LibriCSS \texttt{dev} set (excluding the 0L and OV10 sessions), $K=3$, and $T=15$ in Algorithm~\ref{alg:mixture}.
A reverberated copy of LSMix, named \textbf{LSMix-reverb}, was generated by convolving LSMix with real RIRs collected from the REVERB~\cite{Kinoshita2013TheRC} dataset.
We also added isotropic noises from the REVERB data, and perturbed the loudness between -20 dB and -25 dB using the \texttt{pyloudnorm} tool~\cite{steinmetz2021pyloudnorm}.
We will refer to the combination of LSMix-clean and LSMix-reverb as \textbf{LSMix-full} hereafter.
Ablation experiments were conducted on the \textit{anechoic} LibriCSS by training SURT on LSMix-clean. 
For these experiments, we used on-the-fly noise augmentation using noises from the MUSAN corpus~\cite{Snyder2015MUSANAM}.
For final evaluation, we trained SURT on LSMix-full ($\sim$4400h), so that the same model can be used on both anechoic and replayed LibriCSS.
All models were trained using on-the-fly SpecAugment~\cite{Park2019SpecAugmentAS}.
We found it beneficial to use high overlap mixtures in the warm-up stage of SURT training to encourage better mask estimation~\cite{Boeddeker2023TSSEPJD}. 
%
%
For pre-training on single-speaker data, we used LibriSpeech \texttt{train} set, optionally convolved with synthetic RIRs (for the final evaluation).
This pre-training was done for 10 epochs.

Since AMI and ICSI have similar characteristics, we trained a combined SURT model for them using synthetic mixtures created from close-talk utterances, again using sub-segments obtained from forced alignments ($\tau=0.5$).
These mixtures were obtained by setting $D_{=\text{spk}}$, $D_{\neq\text{spk}}$, $D_{\text{ovl}}$, and $P_{\text{ovl}}$ as 0.5, 0.5, 1.0, and 0.8, respectively, in Algorithm~\ref{alg:mixture}.
$K$ and $T$ were set to 3 and 15s, respectively, similar to LSMix-clean simulation.
We refer to these mixtures as \textbf{AIMix-clean}, their reverberant copy as \textbf{AIMix-reverb}, and the combination as \textbf{AIMix-full}.
The models were subsequently adapted by combining the real train sessions from all microphone settings.
%

\noindent
\textbf{Hyper-parameters.}
The auxiliary loss scales, $\lambda_{\text{ctc}}$ and $\lambda_{\text{mask}}$, were set to 0.2 each.
$\mathcal{L}_{\text{mask}}$ was not used during adaptation since ground-truth separated audio may not be available for real data. 
We trained the models with the ScaledAdam optimizer following the standard zipformer-transducer recipes in \texttt{icefall}.
This is a variant of Adam where each parameter's update is scaled proportional to the norm of that parameter.
The learning rate was warmed up to 0.004 for 5000 iterations, and decayed exponentially thereafter.
All models were trained for 30 epochs using 4 GPUs\footnote{We used either Titan RTX (with batch size 500s) or V100 (with batch size 650s) depending on availability on our compute cluster.}.
%
%

\noindent
\textbf{Decoding.}
Checkpoints from the last 9 epochs were averaged for decoding.
We conducted experiments with both greedy decoding and beam search. 
For the ablation experiments, we used greedy decoding for faster turn-around. 
For the final evaluation, we used a ``modified'' version of beam search with beam size of 4.
This variant constraints the emission of at most 1 non-blank token at each time step, which allows batched decoding. 
We normalized the replayed LibriCSS recordings to -23 dB using \texttt{pyloudnorm} for inference.

\section{Results \& Discussion}
\label{sec:results}

\subsection{Comparison with baselines}
\label{sec:res_baseline}
First, we compared the SURT 2.0 \textit{base} and \textit{large} models with the baselines from literature: SURT~\cite{Raj2021ContinuousSM} and MT-RNNT~\cite{Sklyar2021MultiTurnRF}, on the LibriCSS ``anechoic'' and ``replayed'' recordings.
These comparisons are shown in Table~\ref{tab:results}.
We cannot compare SURT with published results on t-SOT~\cite{Kanda2022StreamingMA} since the latter is evaluated using the \texttt{asclite}-based speaker-agnostic WER (SAg-WER) metric.
SAg-WER requires token-level time-stamp estimation, and does not penalize an utterance being split into multiple ``channels,'' which is penalized by ORC-WER.
The models were trained on LSMix-full, and adapted to LibriCSS using the \texttt{dev} set. 
For adaptation, we segmented the \texttt{dev} sessions in 2 ways, by cutting at maximum pause durations of 0.1s and 0.5s, respectively.
This resulted in 381 sub-sessions of average duration 17.9s, totalling approximately 1.88h each from anechoic and replayed conditions and containing 18.6\% overlapped speech.
The model was then trained on the combined adaptation data for 8 epochs (for \textit{base}) or 15 epochs (for \textit{large}).
The learning rate was warmed up to 0.0004 for 2 (or 4) epochs and decayed thereafter.
A single V100 GPU was used for adaptation.

As the overlap ratio increased from 0\% to 40\%, the WER also increased, which is expected. 
Even without adaptation, SURT-base outperformed the original SURT on the anechoic setting, improving the average WER from 20.0\% to 14.4\%.
This may be because the anechoic training mixtures are well matched to the evaluation condition in the absence of far-field artifacts.
The largest WER difference was on the 0S (0\% overlap with short pauses) condition, where SURT-base achieved 4.7\%, compared to 18.9\% for SURT. 
We attribute this improvement primarily to pre-training on single-speaker data, which allows the model to handle non-overlapping speech well.
On using model adaptation, the anechoic WER further improved by 0.6\% absolute, with consistent improvements across most overlap conditions.
The largest improvement was obtained for the OV40 sessions, where WER reduced from 23.0\% to 20.6\%.
When evaluated on replayed LibriCSS, SURT-base was better than SURT and MT-RNNT on average, but slightly worse on overlapped conditions.
This may be because we used a limited set of real RIRs for simulating reverberant training mixtures, whereas SURT and MT-RNNT used on-the-fly simulated RIRs\footnote{We experimented with using simulated RIRs, but we found that it consistently degraded WERs on the 0S condition, similar to the original SURT.}.
Adaptation on the \texttt{dev} set improved performance across all settings, with the resulting average WER reducing to 20.0\%.
Note that the same model was used to evaluate both settings, whereas in the original SURT, separate models were trained for each setting.

The \textit{large} model followed similar trends as the \textit{base} model, but provided consistent improvements in WER across most conditions.
For unadapted models, larger improvement was observed on the replayed setting compared to the anechoic setting (5.7\% vs. 3.5\% relative).
We conjecture that the larger masking network (6 DP-LSTM layers) may be better suited for unmixing reverberant features, leading to improved WERs.
We also found that the \textit{large} model benefited more from adaptation on in-domain data, perhaps due to higher representation capacity.
The relative WER improvement from adaptation was 12.2\% and 14.2\% for the anechoic and replayed conditions, respectively, whereas for the \textit{base} model, the improvements were 4.2\% and 4.3\%.
Overall, our SURT-large model provided relative WER improvements of 39.0\% (20.0\%~$\rightarrow$~12.2\%) and 26.2\% (22.9\%~$\rightarrow$~16.9\%) over the original SURT.

\subsection{Effect of architectural changes}
\label{sec:results_arch}

\begin{figure*}[b]
\vspace{-2em}
\centering
  \subfloat[Without $\mathcal{L}_{\text{ctc}}$\label{fig:no_ctc_grad}]{%
       \includegraphics[width=0.5\linewidth]{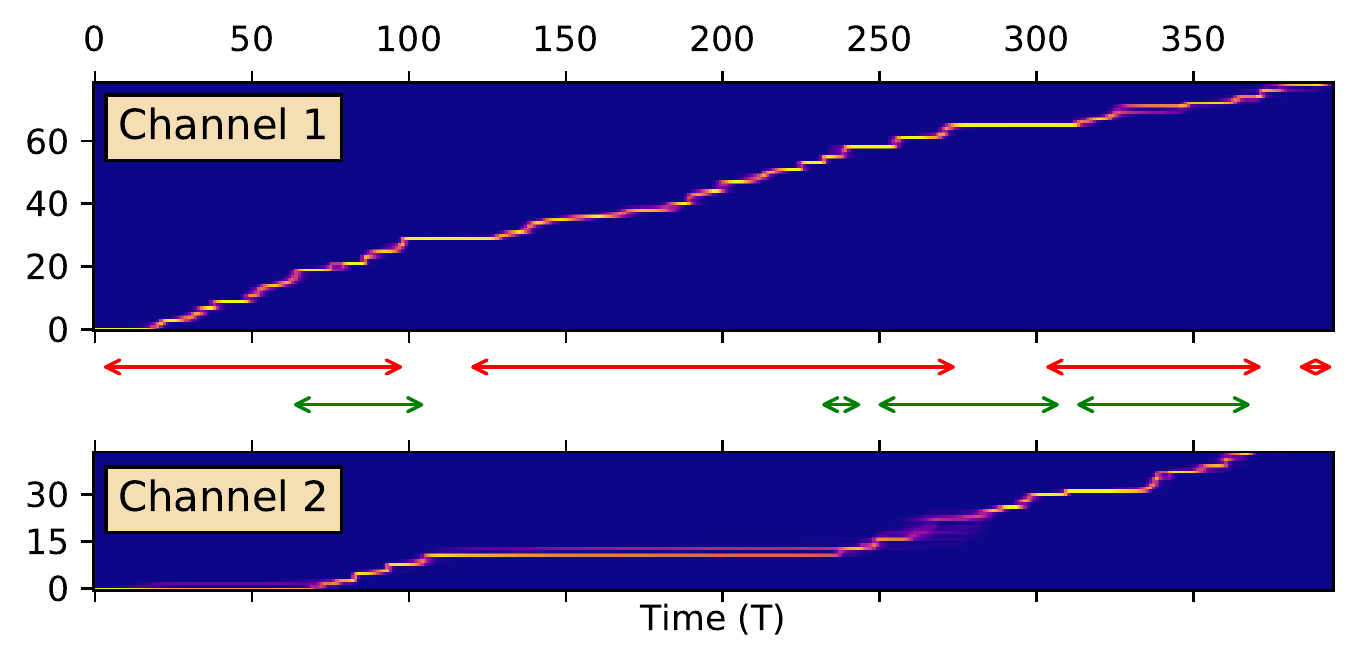}}
    \hfill
  \subfloat[With $\mathcal{L}_{\text{ctc}}$\label{fig:ctc_grad}]{%
        \includegraphics[width=0.5\linewidth]{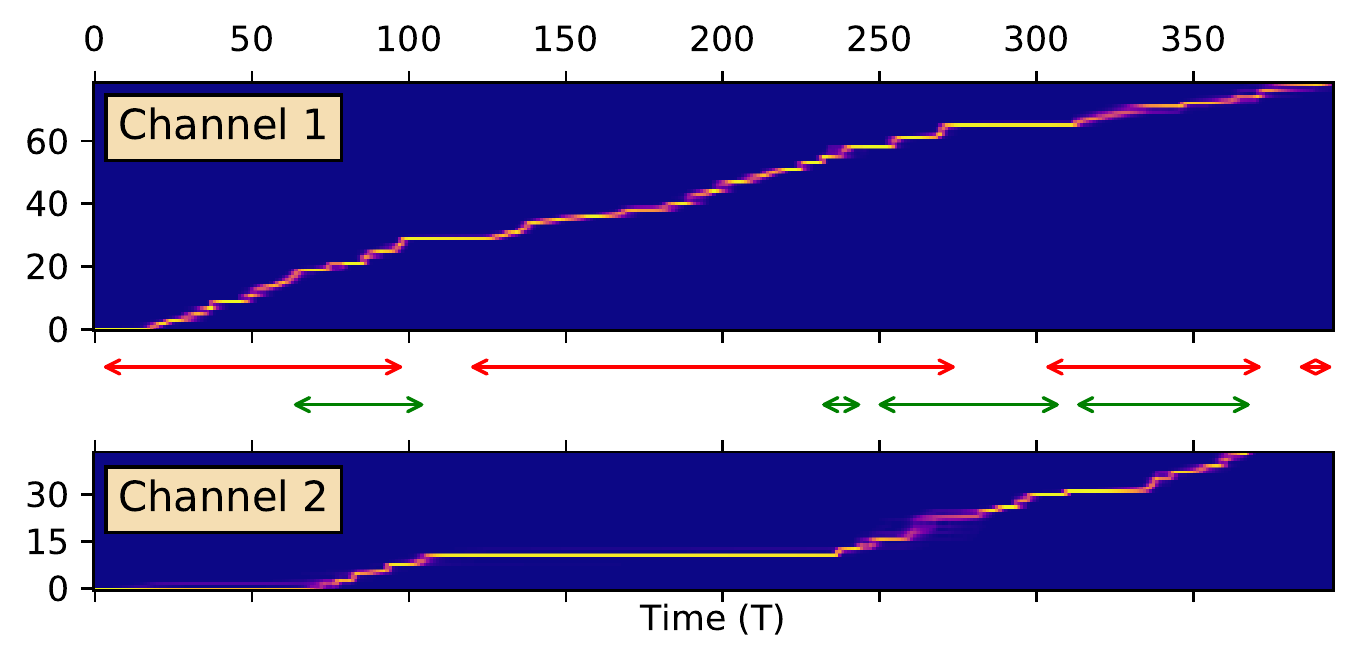}}
\caption{Occupation probabilities of nodes in the $T\times U$ lattice for both output channels, for models trained with and without auxiliary CTC loss ($\mathcal{L}_{\text{ctc}}$. $T$ and $U$ are on $x$ and $y$ axes, respectively. For this mixture, $T=395$, $U_1 = 79$, and $U_2 = 44$. The double arrows between the plots denote the reference segments: red for channel 1 and green for channel 2. Brighter colors denote higher occupation probabilities.}
\label{fig:grad}
\end{figure*}

Recall from Section~\ref{sec:arch} that SURT 2.0 contains several architectural changes compared to the original SURT. 
These include: (i) DP-LSTMs instead of Conv2D in the unmixing module, (ii) branch-tied encoders, and (iii) a stateless prediction network. 
We performed ablation experiments to evaluate the effect of each choice, as shown in Table~\ref{tab:arch}. 
Each of the first three rows denote the performance when one of the components is changed (shown in red), while the last row shows the configuration of the final SURT 2.0 model.
We selected model configurations such that all models have roughly the same number of parameters.
All models were trained on LSMix-clean until convergence,
and evaluated on the anechoic LibriCSS setting with greedy decoding. 
The recognition modules were pre-trained for all setups, but no auxiliary losses or adaptation were used.
We also tried replacing the zipformer encoder with a DP-LSTM (as was used in the original SURT), but this model did not converge.
This may be because the original SURT used 3-fold sub-sampled STFT features as input, and the training mixtures themselves were shorter on average (cf. Section~\ref{sec:sim}).
As a result, the input sequences for the encoder in SURT 2.0 are approximately 3-5 times longer, which may affect convergence in architectures that do not use self-attention.

The largest performance degradation was caused by replacing the DP-LSTM based masking network with Conv2D.
This is consistent with speech separation research where dual-path encoders usually provide large improvements due to their ability to model long sequences~\cite{Luo2019DualPathRE}.
Without the DP-LSTM masking, the unmixing module was effectively futile, as evident by the high WERs on the overlapping sessions.

For the encoder architecture, applying branch tying using equation~(\ref{eq:bt}) provided significant improvements on high overlap conditions.
For example, the relative WER reduction with branch tying was 23.3\% and 19.2\% on the OV30 and OV40 settings, respectively.
The improvement was largely due to a reduction in deletion errors (at the cost of a small increase in insertions).
For example, on the OV40 setting, the deletion error rate reduced from 20.6\% to 14.8\%, with the insertion increasing from 3.5\% to 5.5\%.
This validates our conjecture that branch tying helps in alleviating errors caused by \textit{omissions} where complete segments are skipped by both branches.
We will analyze these errors further in Section~\ref{sec:results_other}.

Finally, we observed small but consistent improvements on replacing the LSTM-based prediction network (used in the original SURT) with a stateless network (i.e., using a Conv1D layer).
The largest relative improvement for this change was seen in the 0S setting, where WER improved by 19.0\% (compared to 7.0\% relative WER improvement overall).
As mentioned in Section~\ref{sec:arch}, we conjecture that this may be because a stateless decoder is more suited to frequent context switching that is required for modeling quick turn-taking.

\begin{table}[t]
\centering
\caption{Effect of architectural choices for various components, shown on ``anechoic'' LibriCSS \texttt{test} set. The last row denotes the final network architectures for SURT 2.0.}
\label{tab:arch}
\adjustbox{max width=\linewidth}{
\begin{tabular}{@{}l@{\hspace{0.1\tabcolsep}}c@{\hspace{0.1\tabcolsep}}ccrrrrrrr@{}}
\toprule
\begin{tabular}{@{}l}\textbf{Masking}\\\textbf{network}\end{tabular} & \begin{tabular}{@{}c}\textbf{Branch}\\\textbf{tying}\end{tabular} &  \begin{tabular}{@{}c}\textbf{Pred.}\\\textbf{network}\end{tabular} & \begin{tabular}{@{}c}\textbf{Size}\\\textbf{(M)}\end{tabular} & \textbf{0L} & \textbf{0S} & \textbf{OV10} & \textbf{OV20} & \textbf{OV30} & \textbf{OV40} & \textbf{Avg.} \\
 \midrule
\textcolor{red}{Conv2D} & \cmark & Conv1D & 25.3 & 13.9 & 6.9 & 27.0 & 35.4 & 41.0 & 45.4 & 28.3 \\
DP-LSTM & \textcolor{red}{\xmark} & Conv1D & 24.6 & 6.6 & 5.4 & 21.3 & 26.6 & 33.1 & 41.2 & 22.4 \\
DP-LSTM & \cmark & \textcolor{red}{LSTM} & 28.1 & 7.6 & 6.3 & \textbf{17.2} & 26.7 & 26.8 & 34.7 & 19.9 \\
\midrule
DP-LSTM & \cmark & Conv1D & 26.7 & \textbf{6.4} & \textbf{5.1} & 17.5 & \textbf{23.5} & \textbf{25.4} & \textbf{33.3} & \textbf{18.5} \\
\bottomrule
\end{tabular}
}
\end{table}

\subsection{Effect of auxiliary objectives}
\label{sec:results_aux}

\begin{table}[t]
\centering
\caption{Effect of auxiliary objectives on ``anechoic'' LibriCSS \texttt{test} set. All models used the SURT-base architecture.}
\label{tab:auxiliary}
\adjustbox{max width=\textwidth}{
\begin{tabular}{@{}ccccccccr@{}}
\toprule
$\mathcal{L}_{\mathrm{ctc}}$ & $\mathcal{L}_{\mathrm{mask}}$ & \textbf{0L} & \textbf{0S} & \textbf{OV10} & \textbf{OV20} & \textbf{OV30} & \textbf{OV40} & \textbf{Avg.} \\ \midrule
\xmark & \xmark & 6.4 & 5.1 & 17.5 & 23.5 & 25.4 & 33.3 & 18.5 \\
\cmark & \xmark & 6.0 & 5.2 & 17.9 & 23.7 & 22.8 & 29.6 & 17.5 \\
\xmark & \cmark & \textbf{5.6} & \textbf{4.9} & 16.3 & 21.3 & 24.6 & 29.5 & 17.1 \\
\cmark & \cmark & 6.1 & 5.0 & \textbf{13.6} & \textbf{19.0} & \textbf{21.1} & \textbf{26.5} & \textbf{15.2} \\
\bottomrule
\end{tabular}
}
\end{table}

We also performed ablation experiments to study the effect of $\mathcal{L}_{\text{ctc}}$ and $\mathcal{L}_{\text{mask}}$. 
Similar to Section~\ref{sec:results_arch}, we trained SURT-base models on LSMix-clean for this investigation, and evaluated the models on anechoic LibriCSS using greedy search. 
The results are shown in Table~\ref{tab:auxiliary}.

First, we see that adding CTC loss on the encoder improved WER mainly for highly overlapped sessions. 
We obtained relative improvements of 10.2\% and 11.1\% on the OV30 and OV40 sessions, respectively. 
For sessions with more overlaps and turn-taking, both output branches may contain several speech segments. 
We conjecture that an auxiliary CTC objective may be useful in aligning the segments to the corresponding audio during training, resulting in better modeling for high overlap sessions. 
To validate our conjecture, we plotted the occupation probabilities for the nodes in the RNN-T lattice (of shape $T \times U$), as obtained from the gradients of the simple additive joiner used in the pruned transducer loss~\cite{Kuang2022PrunedRF}. 
These values should correspond to a soft alignment between the input and the label sequence\footnote{It makes more sense to use this value instead of CTC alignments because (i) the model trained without $\mathcal{L}_{\text{ctc}}$ cannot provide corresponding alignments, and (ii) we use the transducer head for the ASR task.}.
In Fig.~\ref{fig:grad}, we show example plots for a randomly selected mixtures from the training set, using models trained with and without $\mathcal{L}_{\text{ctc}}$.
When the auxiliary CTC loss was used, the model was able to better align the silence region (time frames 100 to 250) in channel 2, as discernible through the bright horizontal line.

Using auxiliary masking loss $\mathcal{L}_{\text{mask}}$, as defined in equation~(\ref{eq:mask}), again improved WER performance over the SURT-base model, as shown in the third row of Table~\ref{tab:auxiliary}.
Surprisingly, we observed most improvements in the low-overlap conditions --- for instance, 12.5\% relative WER reduction for 0L.
Most of this improvement again resulted from reduction in deletion errors (2.0\%~$\rightarrow$~1.1\% for 0L).
We also experimented with using a graph-PIT based masking loss~\cite{vonNeumann2021GraphPITGP} instead of the HEAT-based loss, but it did not provide similar improvements.
This may be because our final transducer objective uses the HEAT formulation.
Finally, the best WER results were obtained on combining both the auxiliary objectives.
The resulting model demonstrated a relative WER reduction of 17.8\% over the SURT-base model trained without these objectives.

\begin{figure}[t]
\centering
\includegraphics[width=0.7\linewidth]{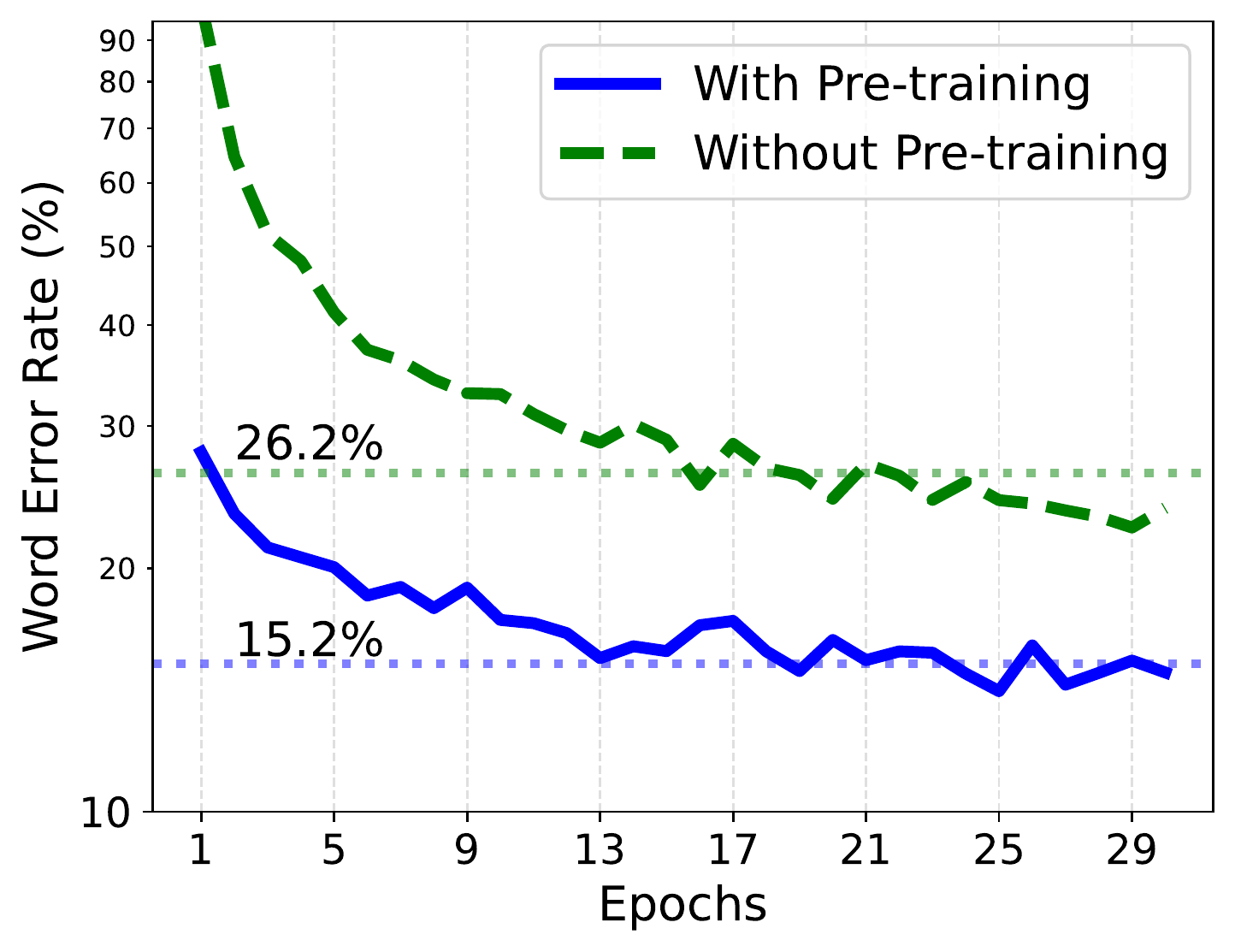}
\vspace{-0.5em}
\caption{Effect of transducer pre-training with LibriSpeech. WERs are shown for the LibriCSS \texttt{dev} set after each training epoch. The dotted horizontal lines show the final WERs on \texttt{test} set with model averaging.}
\label{fig:pretrain}
\end{figure}

\begin{figure*}[t]
\centering
  \subfloat[\textbf{Anechoic}\label{fig:anechoic_beam}]{%
       \includegraphics[width=0.5\linewidth]{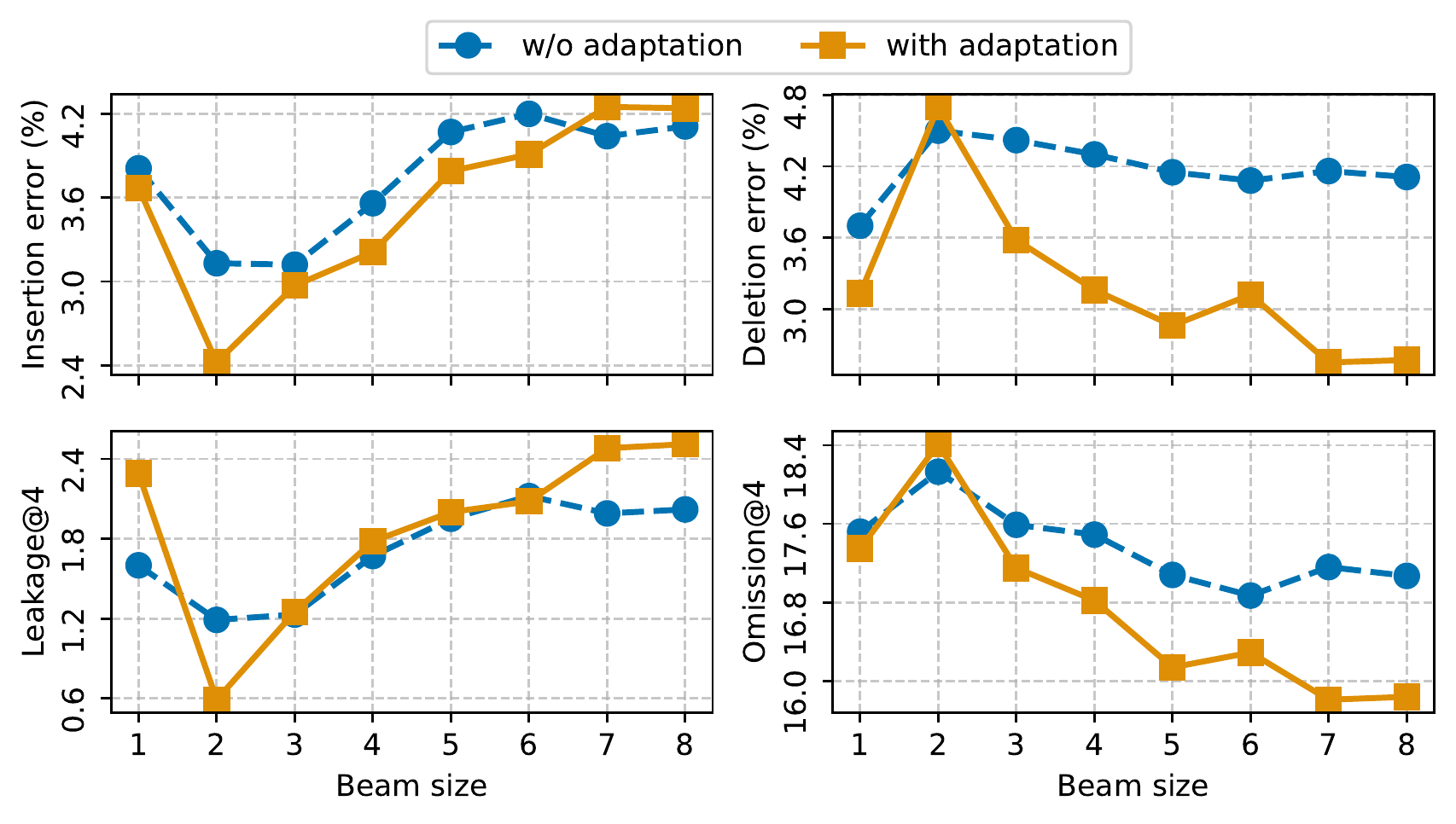}}
    \hfill
  \subfloat[\textbf{Replayed}\label{fig:replayed_beam}]{%
        \includegraphics[width=0.5\linewidth]{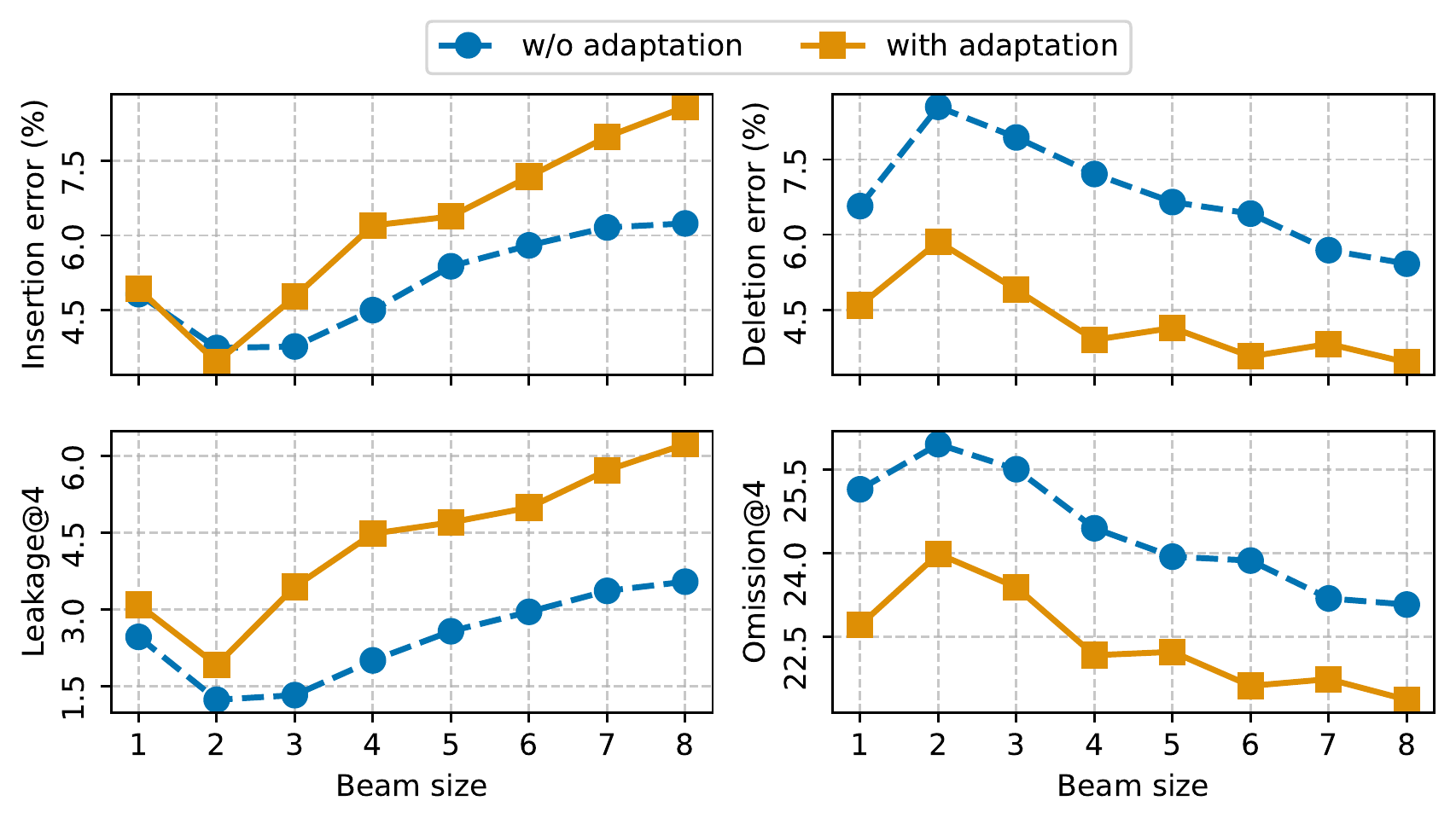}}
\vspace{-0.5em}
\caption{Effect of decoding beam size on insertion and deletion errors, versus effect on leakage and omission (for $n=4$), for SURT-base models with and without adaptation, for (a) ``anechoic'' and (b) ``replayed'' conditions, averaged across all overlap settings. The top row shows insertion and deletion errors, while the bottom row contains leakage@4 and omission@4.}
\label{fig:beams}
\vspace{-1em}
\end{figure*}

\subsection{Effect of pre-training}
\label{sec:results_pretrain}

Pre-training on single-speaker utterances was found to be one of the most effective strategies for faster and better convergence of SURT models\footnote{Such a pre-training strategy has also been used in recent work on multi-channel serialized output training~\cite{Kanda2022VarArrayMT}.}.
To quantify this improvement, we computed the WER on the anechoic LibriCSS \texttt{dev} set (averaged across all overlap conditions) after each epoch of training a SURT-base model on the LSMix-clean data, as shown in Fig.~\ref{fig:pretrain}.
We observe that when pre-training was used, the SURT model converged much faster and to a better WER.
In fact, it surpassed the model without pre-training after just 5 epochs.
On the anechoic \texttt{test} set, the models obtained WER of 26.2\% and 15.2\%, respectively, as shown by the dotted horizontal lines.
The increase in computational time is marginal, since we only pre-train the transducer for 10 epochs (instead of training to convergence).
Alternatively, an off-the-shelf streaming transducer may also be plugged in for this purpose, since the \textit{branch tying} of the encoder is only added at the time of SURT training.
Such a pre-training scheme is analogous to a curriculum learning strategy using single-speaker utterances, with the masks fixed as $\mathbf{M}_1 = J_{T,F}$ and $\mathbf{M}_c = 0 \cdot J_{T,F}$, $\forall c \neq 1$, where $J_{t,f}=1$, $\forall t,f$.

\subsection{Measuring leakage and omission}
\label{sec:results_other}


Throughout this paper, we have mentioned \textit{leakage} and \textit{omission}, first identified in \cite{Raj2021ContinuousSM}, as major contributors of errors in SURT.
In this section, we provide a metric for quantifying these error sources, in terms of n-gram counts on the reference $\mathbf{Y}$ and hypotheses $\hat{\mathbf{Y}}$, as defined in Section~\ref{sec:mt-asr}.

\begin{description}
    \item[omission@n] For some $n$, the fraction of all unique n-grams in $\mathbf{Y}$ not present in any $\hat{\mathbf{Y}}_c$.
    \item[leakage@n] For some $n$, the fraction of all unique n-grams in $\mathbf{Y}$ present in multiple $\hat{\mathbf{Y}}_c$.
\end{description}

If emission time-stamps are known, these definitions may be modified to include time windows for n-gram search.
In their absence, the leakage@n and omission@n rates are upper and lower bounds on the actual leakage and omission, respectively.
Nevertheless, by quantifying these error types explicitly, we can gain some insights about the model behavior.
Our SURT model in this paper does not predict token time-stamps, so we estimate omission and leakage over the entire hypotheses.

In Table~\ref{tab:results}, we showed WERs achieved by the SURT models with and without adaptation, when decoded using beam search with a beam of size 4.
In general, ASR model performance improves on increasing the beam size, but we found that for SURT models, the WER first improved (up to a beam size of 4) and then degraded.
This trend can be explained by looking at the leakage and omission errors, as shown in Fig.~\ref{fig:beams}.
We used the SURT-base model (with and without adaptation) for decoding the anechoic and replayed sets, using beam sizes varying from 1 to 8.
In the figure, we show insertion and deletion error rates, as well as leakage@4 and omission@4.
We observe that leakage first decreased (from beam size 1 to 2) but then increased gradually, while the opposite trend was observed for omission.
Furthermore, adapted models reduce omissions significantly, at the cost of increase in leakage (particularly in the replayed setting).

\begin{table}[t]
\setlength{\tabcolsep}{5pt}
\centering
\caption{Comparison of leakage@4 and omission@4 with insertion and deletion errors on the ``anechoic'' LibriCSS \texttt{dev} set.}
\vspace{-0.5em}
\label{tab:aux2}
\adjustbox{max width=0.7\textwidth}{
\begin{tabular}{@{}ccrrrr@{}}
\toprule
$\mathcal{L}_{\mathrm{ctc}}$ & $\mathcal{L}_{\mathrm{mask}}$ & \textbf{Ins.} & \textbf{Del.} & \textbf{L@4} & \textbf{O@4} \\ \midrule
\xmark & \xmark & 3.44 & 5.96 & 2.44 & 20.63 \\
\cmark & \xmark & 2.56 & 4.87 & 1.51 & 20.10 \\
\xmark & \cmark & 2.96 & 5.88 & 1.49 & 20.34 \\
\cmark & \cmark & 3.03 & 4.65 & 1.68 & 19.27 \\
\bottomrule
\end{tabular}
}
\end{table}

In the above analysis, the overall trend for leakage and omissions followed those of insertion and deletion errors, which is expected.
However, this may not always be the case.
In Table~\ref{tab:auxiliary}, we showed results for ablation experiments done to investigate the effect of auxiliary objectives.
We calculated the leakage@4 and omission@4 rates for those models on the anechoic \texttt{dev} set, and compared them with the corresponding insertion and deletion rates.
The comparison is shown in Table~\ref{tab:aux2}.
We see that although the models trained with $\mathcal{L}_{\text{ctc}}$ and $\mathcal{L}_{\text{mask}}$ have similar L@4 and O@4, the difference in insertion and deletion errors is comparatively large.
This suggests that the additional insertions or deletions are not caused due to unmixing errors, and are most likely due to errors in the recognition module.
This hypothesis seems reasonable because $\mathcal{L}_{\text{ctc}}$ should be more effective than $\mathcal{L}_{\text{mask}}$ in reducing purely ASR-related errors.
When we further include $\mathcal{L}_{\text{mask}}$ in training (last row), both O@4 and deletion reduce by roughly 5\% relative, indicating that the improvement almost entirely results from recovered sub-segments.

\subsection{Results on AMI and ICSI}
\label{sec:results_ami}

Finally, we evaluated our SURT models on two meeting benchmarks, AMI and ICSI, to verify their efficacy on real-world settings. 
For these experiments, we initialized SURT \textit{base} and \textit{large} models with the final checkpoint from the LSMix-full training, and continued training on the AIMix-full simulated mixtures for 30 epochs. 
The models were further adapted on the AMI and ICSI \texttt{train} sessions, by combining the IHM-Mix, SDM, and beamformed MDM recordings. 
For the adaptation, we cut the sessions at pauses of 0.0 and 0.5 seconds (thus creating two copies with different segmentations).
We trimmed the long sessions to approximately 30s each based on utterance end time marks. 
This process resulted in 330k train samples of average duration 6.4s, totaling 590h and 18.4\% overlapped speech.
We adapted the SURT models on these sub-sessions by training with a LR of 0.0001 for 20 epochs, and used the final checkpoint for inference. 
The results with and without in-domain adaptation are shown in Table~\ref{tab:ami}.

We found that the \textit{large} model obtained consistently better WERs than the \textit{base} model, which is expected. 
All models obtained lower WERs on ICSI, which may be due to its lower overlapped speech ratio (13.6\%) compared to AMI (21.0\%). 
A similar increase in WERs with higher overlaps was also observed for LibriCSS. 
Without model adaptation, the SURT models obtained reasonable WERs for the IHM-Mix and beamformed MDM settings, but not for SDM. 
For instance, the SURT-base model obtained 64.3\% relatively worse WERs on SDM, as compared to IHM-Mix. 
This suggests that purely simulated mixtures cannot compensate for real, far-field training data. 
SDM performance improved significantly with in-domain adaptation, with relative WER decrease of 28.3\% and 43.5\% for SURT-base on AMI and ICSI, respectively.

\begin{table}[t]
\centering
\caption{Results on the AMI and ICSI \texttt{test} sets, under different microphone settings, using SURT \textit{base} and \textit{large} models.}
\vspace{-0.5em}
\label{tab:ami}
\adjustbox{max width=0.9\linewidth}{
\begin{tabular}{@{}lc@{\hspace{2\tabcolsep}}ccc@{\hspace{2\tabcolsep}}cc@{}}
\toprule
\multirow{2}{*}{\textbf{Model}} & \multirow{2}{*}{\textbf{Adapt.}} & \multicolumn{3}{c}{\textbf{AMI}} & \multicolumn{2}{c}{\textbf{ICSI}} \\
\cmidrule(l{1pt}r){3-5} \cmidrule(l{1pt}r{1pt}){6-7}
 & & IHM-Mix & SDM & MDM & IHM-Mix & SDM \\
\midrule
\multirow{2}{*}{\textbf{Base}} & \xmark & 39.8 & 65.4 & 46.6 & 28.3 & 60.0 \\
 & \cmark & 37.4 & 46.9 & 43.7 & 26.3 & 33.9 \\
\midrule
\multirow{2}{*}{\textbf{Large}} & \xmark & 36.8 & 62.5 & 44.4 & 27.8 & 59.7 \\
 & \cmark & 35.1 & 44.6 & 41.4 & 24.4 & 32.2 \\
\bottomrule
\end{tabular}
}
\end{table}

Error analysis revealed that most of the errors were caused by deletion, as shown in Table~\ref{tab:wer_breakdown}. 
For the unadapted models, deletion comprised 89.4\% and 83.2\% of the overall errors for the base and large SURT variants, respectively. 
We attribute this primarily to the model missing very short utterances and back-channels such as ``Okay,'' ``Hmm,'', and so on, which form a significant fraction of such meetings, and which may be useful for downstream dialog understanding tasks.
When we adapted the models using real, in-domain data, the deletion errors reduced significantly, with minor increase in insertion and substitution.
For the adapted models, deletion comprised 68.9\% and 66.8\% of the total errors for the base and large SURT, respectively.
In future work, it may be interesting to investigate models and training objectives which avoid suppression of short overlapping segments.

\begin{table}[t]
\centering
\caption{WER breakdown on the AMI \texttt{test} set for the SDM microphone condition, using SURT \textit{base} and \textit{large} models.}
\label{tab:wer_breakdown}
\vspace{-0.5em}
\adjustbox{max width=0.54\linewidth}{
\begin{tabular}{@{}lccccc@{}}
\toprule
\textbf{Model} & \textbf{Adapt.} & \textbf{Ins.} & \textbf{Del.} & \textbf{Sub.} & \textbf{WER} \\
\midrule
\multirow{2}{*}{\textbf{Base}} & \xmark & 0.8 & 55.8 & 8.8 & 65.4 \\
 & \cmark & 1.8 & 32.3 & 12.8 & 46.9 \\
\midrule
\multirow{2}{*}{\textbf{Large}} & \xmark & 0.9 & 52.0 & 9.7 & 62.5 \\
 & \cmark & 2.1 & 29.8 & 12.8 & 44.6 \\
\bottomrule
\end{tabular}
}
\end{table}

\section{Conclusions and Future Work}
\label{sec:conclusion}

We performed a detailed investigation of the SURT model for multi-talker speech recognition.
By decomposing the continuous, streaming, multi-talker ASR problem into the three components of sparsely overlapped speech separation, long-form ASR, and quick turn-taking modeling, we were able to identify model design strategies that improve performance on one or more of these sub-problems.
We modified SURT's unmixing component to perform mask estimation on filter-bank inputs, which allowed the use of transducer pre-training on single-speaker utterances.
To improve training efficiency, we applied zipformer blocks in the encoder which aggressively subsample the input sequence and use shared attention masks within the blocks.
We also used sub-segments instead of full utterances to simulate training mixtures, which resulted in more frequent turn-taking without increasing the training sequence length.
We further replaced the full-sum transducer loss with the recently proposed pruned transducer to reduce memory requirement for loss computation.
We found that using auxiliary objectives for the encoder and the masking network also improves the model's performance, for instance by producing better soft alignments of the input and output sequences during training, or by reducing leakage in single-speaker regions.
To further reduce errors caused by leakage and omission, we used dual-path LSTMs instead of convolutional layers in unmixing, and added branch tying of encoder outputs in the recognition component.
We trained SURT 2.0 in multiple stages: (i) single-speaker pre-training, (ii) training on simulated mixtures, and (iii) adaptation on in-domain real data, to outperform the larger and computationally expensive SURT models proposed previously on LibriCSS.
We also demonstrated the viability of these models for real meeting benchmarks, namely AMI and ICSI.

Although SURT has shown promising results while being efficient for training and inference, there are several avenues for further enhancements.
First, our quantification of omission and leakage related errors for different beam sizes suggests that specialized decoding methods may be required which process all output branches simultaneously.
For this, we can either extended the prefix beam search technique to process all output branches, or leverage HMM-based joint decoding techniques such as those proposed in \cite{Kocour2021RevisitingJD}.
In order to perform joint speaker diarization with SURT, we can predict word-level time-stamps and speaker change tokens, similar to ideas explored in~\cite{Xia2021TurntoDiarizeOS}.
Another interesting direction to exploit full dialog context may be cross-channel conversational rescoring using large language models~\cite{Xu2022RescoreBERTDS}.
We hope that our open-source implementation of SURT would encourage exploration along one or more of these directions.


\setstretch{0.9}
{\small
\section*{Acknowledgments}
We thank Aparna Khare for discussions and insights on multi-turn RNN-T, and Piotr Zelasko, Fangjun Kuang, and Thilo von Neumann for help with Lhotse, Icefall, and \texttt{meeteval}, respectively. We are grateful to the anonymous reviewers for their feedback.
}

\bibliographystyle{IEEEtran}
\bibliography{main.bib}

\begin{thebibliography}{100}
\providecommand{\url}[1]{#1}
\csname url@samestyle\endcsname
\providecommand{\newblock}{\relax}
\providecommand{\bibinfo}[2]{#2}
\providecommand{\BIBentrySTDinterwordspacing}{\spaceskip=0pt\relax}
\providecommand{\BIBentryALTinterwordstretchfactor}{4}
\providecommand{\BIBentryALTinterwordspacing}{\spaceskip=\fontdimen2\font plus
\BIBentryALTinterwordstretchfactor\fontdimen3\font minus
  \fontdimen4\font\relax}
\providecommand{\BIBforeignlanguage}[2]{{%
\expandafter\ifx\csname l@#1\endcsname\relax
\typeout{** WARNING: IEEEtran.bst: No hyphenation pattern has been}%
\typeout{** loaded for the language `#1'. Using the pattern for}%
\typeout{** the default language instead.}%
\else
\language=\csname l@#1\endcsname
\fi
#2}}
\providecommand{\BIBdecl}{\relax}
\BIBdecl

\bibitem{Amodei2016DS2}
D.~Amodei \emph{et~al.}, ``Deep speech 2 : End-to-end speech recognition in
  english and mandarin,'' in \emph{ICML}, 2016.

\bibitem{Xiong2017TowardHP}
W.~Xiong, J.~Droppo, X.~Huang, F.~Seide, M.~Seltzer, A.~Stolcke, D.~Yu, and
  G.~Zweig, ``Toward human parity in conversational speech recognition,''
  \emph{IEEE/ACM TASLP}, vol.~25, pp. 2410--2423, 2017.

\bibitem{li2021recent}
J.~Li, ``Recent advances in end-to-end automatic speech recognition,''
  \emph{APSIPA Transactions on Signal and Information Processing}, 2021.

\bibitem{Barker2015TheT}
J.~Barker, R.~Marxer, E.~Vincent, and S.~Watanabe, ``The third {‘CHiME’}
  speech separation and recognition challenge: Dataset, task and baselines,''
  in \emph{IEEE ASRU}, 2015.

\bibitem{Kinoshita2013TheRC}
K.~Kinoshita, M.~Delcroix, T.~Yoshioka, T.~Nakatani, A.~Sehr, W.~Kellermann,
  and R.~Maas, ``The {REVERB} challenge: A common evaluation framework for
  dereverberation and recognition of reverberant speech,'' in \emph{IEEE
  WASPAA}, 2013.

\bibitem{Watanabe2020CHiME6CT}
S.~Watanabe, M.~Mandel, J.~Barker, and E.~Vincent, ``{CHiME-6} challenge:
  Tackling multispeaker speech recognition for unsegmented recordings,''
  \emph{ArXiv}, 2020.

\bibitem{Carletta2005TheAM}
J.~Carletta \emph{et~al.}, ``The {AMI} meeting corpus: A pre-announcement,'' in
  \emph{MLMI}, 2005.

\bibitem{Shriberg2001ObservationsOO}
E.~Shriberg, A.~Stolcke, and D.~Baron, ``Observations on overlap: findings and
  implications for automatic processing of multi-party conversation,'' in
  \emph{InterSpeech}, 2001.

\bibitem{Yoshioka2019MeetingTU}
T.~Yoshioka, D.~Dimitriadis, A.~Stolcke, W.~Hinthorn, Z.~Chen, M.~Zeng, and
  X.~Huang, ``Meeting transcription using asynchronous distant microphones,''
  in \emph{InterSpeech}, 2019.

\bibitem{Fiscus2007TheRT}
J.~G. Fiscus, J.~Ajot, and J.~S. Garofolo, ``The rich transcription 2007
  meeting recognition evaluation,'' in \emph{CLEaR}, 2007.

\bibitem{Hain2012TranscribingMW}
T.~Hain, L.~Burget \emph{et~al.}, ``Transcribing meetings with the {AMIDA}
  systems,'' \emph{IEEE Transactions on Audio, Speech, and Language
  Processing}, vol.~20, pp. 486--498, 2012.

\bibitem{Segbroeck2019DiPCoD}
M.~V. Segbroeck, A.~F.~A. Zaid, K.~Kutsenko, C.~Huerta, T.~Nguyen, X.~Luo,
  B.~Hoffmeister, J.~Trmal, M.~Omologo, and R.~Maas, ``{DiPCo} - dinner party
  corpus,'' in \emph{InterSpeech}, 2019.

\bibitem{Wang2017SupervisedSS}
D.~Wang and J.~Chen, ``Supervised speech separation based on deep learning: An
  overview,'' \emph{IEEE/ACM TASLP}, vol.~26, pp. 1702--1726, 2017.

\bibitem{Raj2020IntegrationOS}
D.~Raj, P.~Denisov \emph{et~al.}, ``Integration of speech separation,
  diarization, and recognition for multi-speaker meetings: System description,
  comparison, and analysis,'' in \emph{IEEE SLT}, 2021.

\bibitem{Wu2021InvestigationOP}
J.~Wu, Z.~Chen, S.~Chen, Y.~Wu, T.~Yoshioka, N.~Kanda, S.~Liu, and J.~Li,
  ``Investigation of practical aspects of single channel speech separation for
  {ASR},'' in \emph{InterSpeech}, 2021.

\bibitem{Yu2017RecognizingMS}
D.~Yu, X.~Chang, and Y.~Qian, ``Recognizing multi-talker speech with
  permutation invariant training,'' in \emph{InterSpeech}, 2017.

\bibitem{Qian2017SingleChannelMS}
Y.~Qian, X.~Chang, and D.~Yu, ``Single-channel multi-talker speech recognition
  with permutation invariant training,'' \emph{Speech Communication}, vol. 104,
  pp. 1--11, 2017.

\bibitem{seki-etal-2018-purely}
H.~Seki, T.~Hori, S.~Watanabe, J.~Le~Roux, and J.~R. Hershey, ``A purely
  end-to-end system for multi-speaker speech recognition,'' in \emph{ACL},
  2018.

\bibitem{Kanda2020SerializedOT}
N.~Kanda, Y.~Gaur, X.~Wang, Z.~Meng, and T.~Yoshioka, ``Serialized output
  training for end-to-end overlapped speech recognition,'' in
  \emph{InterSpeech}, 2020.

\bibitem{Graves2012SequenceTW}
A.~Graves, ``Sequence transduction with recurrent neural networks,'' in
  \emph{ICML Representation Learning Workshop}, 2012.

\bibitem{he2019streaming}
Y.~He, T.~N. Sainath, R.~Prabhavalkar, I.~McGraw, R.~Alvarez, D.~Zhao,
  D.~Rybach, A.~Kannan, Y.~Wu, R.~Pang \emph{et~al.}, ``Streaming end-to-end
  speech recognition for mobile devices,'' in \emph{IEEE ICASSP}, 2019.

\bibitem{Wu2020StreamingTA}
C.~Wu, Y.~Wang, Y.~Shi, C.-F. Yeh, and F.~Zhang, ``{Streaming Transformer-Based
  Acoustic Models Using Self-Attention with Augmented Memory},'' in
  \emph{InterSpeech}, 2020.

\bibitem{Li2019RNNT}
J.~Li, R.~Zhao, H.~Hu, and Y.~Gong, ``Improving {RNN} transducer modeling for
  end-to-end speech recognition,'' in \emph{IEEE ASRU}, 2019.

\bibitem{Kanda2022StreamingMA}
N.~Kanda, J.~Wu, Y.~Wu, X.~Xiao, Z.~Meng, X.~Wang, Y.~Gaur, Z.~Chen, J.~Li, and
  T.~Yoshioka, ``Streaming multi-talker {ASR} with token-level serialized
  output training,'' in \emph{InterSpeech}, 2022.

\bibitem{Lu2020StreamingEM}
L.~Lu, N.~Kanda, J.~Li, and Y.~Gong, ``Streaming end-to-end multi-talker speech
  recognition,'' \emph{IEEE Signal Processing Letters}, vol.~28, pp. 803--807,
  2020.

\bibitem{Sklyar2021StreamingMA}
I.~Sklyar, A.~Piunova, and Y.~Liu, ``Streaming multi-speaker {ASR} with
  {RNN-T},'' in \emph{IEEE ICASSP}, 2021.

\bibitem{Raj2021ContinuousSM}
D.~Raj, L.~Lu, Z.~Chen, Y.~Gaur, and J.~Li, ``Continuous streaming multi-talker
  {ASR} with dual-path transducers,'' in \emph{IEEE ICASSP}, 2022.

\bibitem{Sklyar2021MultiTurnRF}
I.~Sklyar, A.~Piunova, X.~Zheng, and Y.~Liu, ``Multi-turn {RNN-T} for streaming
  recognition of multi-party speech,'' in \emph{IEEE ICASSP}, 2021.

\bibitem{Lu2021StreamingMS}
L.~Lu, N.~Kanda, J.~Li, and Y.~Gong, ``Streaming multi-talker speech
  recognition with joint speaker identification,'' in \emph{InterSpeech}, 2021.

\bibitem{Lu2022EndpointDF}
L.~Lu, J.~Li, and Y.~Gong, ``Endpoint detection for streaming end-to-end
  multi-talker {ASR},'' in \emph{IEEE ICASSP}, 2022.

\bibitem{Sklyar2022SeparatorTransducerSegmenterSR}
I.~Sklyar, A.~Piunova, and C.~Osendorfer, ``Separator-transducer-segmenter:
  Streaming recognition and segmentation of multi-party speech,'' in
  \emph{InterSpeech}, 2022.

\bibitem{Chang2018MonauralMS}
X.~Chang, Y.~Qian, and D.~Yu, ``Monaural multi-talker speech recognition with
  attention mechanism and gated convolutional networks,'' in
  \emph{InterSpeech}, 2018.

\bibitem{Chang2018AdaptivePI}
------, ``Adaptive permutation invariant training with auxiliary information
  for monaural multi-talker speech recognition,'' in \emph{IEEE ICASSP}, 2018.

\bibitem{Tan2018KnowledgeTI}
T.~Tan, Y.~Qian, and D.~Yu, ``Knowledge transfer in permutation invariant
  training for single-channel multi-talker speech recognition,'' in \emph{IEEE
  ICASSP}, 2018.

\bibitem{Graves2006ConnectionistTC}
A.~Graves, S.~Fern{\'a}ndez, F.~Gomez, and J.~Schmidhuber, ``Connectionist
  temporal classification: labelling unsegmented sequence data with recurrent
  neural networks,'' in \emph{ICML}, 2006.

\bibitem{Lu2016OnTT}
L.~Lu, X.~Zhang, and S.~Renals, ``On training the recurrent neural network
  encoder-decoder for large vocabulary end-to-end speech recognition,'' in
  \emph{IEEE ICASSP}, 2016.

\bibitem{Chorowski2015AttentionBasedMF}
J.~Chorowski, D.~Bahdanau, D.~Serdyuk, K.~Cho, and Y.~Bengio, ``Attention-based
  models for speech recognition,'' in \emph{NIPS}, 2015.

\bibitem{Chiu2018StateoftheArtSR}
C.~Chiu, T.~Sainath \emph{et~al.}, ``State-of-the-art speech recognition with
  sequence-to-sequence models,'' in \emph{IEEE ICASSP}, 2018.

\bibitem{Chang2020EndToEndMS}
X.~Chang, W.~Zhang, Y.~Qian, J.~L. Roux, and S.~Watanabe, ``End-to-end
  multi-speaker speech recognition with transformer,'' in \emph{IEEE ICASSP},
  2020.

\bibitem{Denisov2019EndtoEndMS}
P.~Denisov and N.~T. Vu, ``End-to-end multi-speaker speech recognition using
  speaker embeddings and transfer learning,'' in \emph{InterSpeech}, 2019.

\bibitem{Zhang2020ImprovingES}
W.~Zhang, X.~Chang, Y.~Qian, and S.~Watanabe, ``Improving end-to-end
  single-channel multi-talker speech recognition,'' \emph{IEEE/ACM Transactions
  on Audio, Speech, and Language Processing}, vol.~28, pp. 1385--1394, 2020.

\bibitem{Lin2022SeparatetoRecognizeJM}
Y.~Lin, Z.~Du, S.~Zhang, F.~Yu, Z.~Zhao, and F.~Wu, ``Separate-to-recognize:
  Joint multi-target speech separation and speech recognition for
  speaker-attributed {ASR},'' in \emph{ISCSLP}, 2022.

\bibitem{Chang2019EndtoendMM}
X.~Chang, Y.~Qian, K.~Yu, and S.~Watanabe, ``End-to-end monaural multi-speaker
  {ASR} system without pretraining,'' in \emph{IEEE ICASSP}, 2019.

\bibitem{Shi2022TrainFS}
J.~Shi, X.~Chang, S.~Watanabe, and B.~Xu, ``Train from scratch: Single-stage
  joint training of speech separation and recognition,'' \emph{Computer,
  Speech, and Language}, vol.~76, p. 101387, 2022.

\bibitem{vonNeumann2019EndtoEndTO}
T.~von Neumann, K.~Kinoshita, L.~Drude, C.~Boeddeker, M.~Delcroix, T.~Nakatani,
  and R.~Haeb-Umbach, ``End-to-end training of time domain audio separation and
  recognition,'' in \emph{IEEE ICASSP}, 2019.

\bibitem{Inaguma2020EnhancingMM}
H.~Inaguma, M.~Mimura, and T.~Kawahara, ``Enhancing monotonic multihead
  attention for streaming {ASR},'' in \emph{InterSpeech}, 2020.

\bibitem{Li2022TransformerBasedSA}
M.~Li, S.~Zhang, C.~Zorila, and R.~Doddipatla, ``Transformer-based streaming
  {ASR} with cumulative attention,'' in \emph{IEEE ICASSP}, 2022.

\bibitem{Tsunoo2019TransformerAW}
E.~Tsunoo, Y.~Kashiwagi, T.~Kumakura, and S.~Watanabe, ``Transformer {ASR} with
  contextual block processing,'' in \emph{IEEE ASRU}, 2019.

\bibitem{Isik2016SingleChannelMS}
Y.~Z. Isik, J.~L. Roux, Z.~Chen, S.~Watanabe, and J.~R. Hershey,
  ``Single-channel multi-speaker separation using deep clustering,'' in
  \emph{InterSpeech}, 2016.

\bibitem{Kanda2020JointSC}
N.~Kanda, Y.~Gaur \emph{et~al.}, ``Joint speaker counting, speech recognition,
  and speaker identification for overlapped speech of any number of speakers,''
  in \emph{InterSpeech}, 2020.

\bibitem{Kanda2020InvestigationOE}
N.~Kanda, X.~Chang, Y.~Gaur, X.~Wang, Z.~Meng, Z.~Chen, and T.~Yoshioka,
  ``Investigation of end-to-end speaker-attributed {ASR} for continuous
  multi-talker recordings,'' in \emph{IEEE SLT}, 2020.

\bibitem{Kanda2022StreamingSA}
N.~Kanda, J.~Wu, Y.~Wu, X.~Xiao, Z.~Meng, X.~Wang, Y.~Gaur, Z.~Chen, J.~Li, and
  T.~Yoshioka, ``Streaming speaker-attributed {ASR} with token-level speaker
  embeddings,'' in \emph{InterSpeech}, 2022.

\bibitem{Kanda2022VarArrayMT}
N.~Kanda, J.~Wu, X.~Wang, Z.~Chen, J.~Li, and T.~Yoshioka, ``Vararray meets
  {t-SOT}: Advancing the state of the art of streaming distant conversational
  speech recognition,'' in \emph{IEEE ICASSP}, 2023.

\bibitem{Kanda2021LargeScalePO}
N.~Kanda, G.~Ye, Y.~Wu, Y.~Gaur, X.~Wang, Z.~Meng, Z.~Chen, and T.~Yoshioka,
  ``Large-scale pre-training of end-to-end multi-talker {ASR} for meeting
  transcription with single distant microphone,'' in \emph{InterSpeech}, 2021.

\bibitem{Yu2021M2MetTI}
F.~Yu, S.~Zhang, Y.~Fu, L.~Xie, S.~Zheng, Z.~Du, W.~Huang, P.~Guo, Z.~Yan,
  B.~Ma, X.~Xu, and H.~Bu, ``{M2MeT}: The {ICASSP} 2022 multi-channel
  multi-party meeting transcription challenge,'' in \emph{IEEE ICASSP}, 2021.

\bibitem{Yu2022SummaryOT}
F.~Yu, S.~Zhang, P.~Guo, Y.~Fu, Z.~Du, S.~Zheng, W.~Huang, L.~Xie, Z.~Tan,
  D.~Wang, Y.~Qian, K.-A. Lee, Z.~Yan, B.~Ma, X.~Xu, and H.~Bu, ``Summary on
  the {ICASSP} 2022 multi-channel multi-party meeting transcription grand
  challenge,'' in \emph{IEEE ICASSP}, 2022.

\bibitem{Chen2020ContinuousSS}
Z.~Chen, T.~Yoshioka, L.~Lu, T.~Zhou, Z.~Meng, Y.~Luo, J.~Wu, and J.~Li,
  ``Continuous speech separation: Dataset and analysis,'' in \emph{IEEE
  ICASSP}, 2020.

\bibitem{Wu2020AnEA}
J.~Wu, Z.~Chen, J.~Li, T.~Yoshioka, Z.~Tan, E.~Lin, Y.~Luo, and L.~Xie, ``An
  end-to-end architecture of online multi-channel speech separation,'' in
  \emph{InterSpeech}, 2020.

\bibitem{Yoshioka2018RecognizingOS}
T.~Yoshioka, H.~Erdogan, Z.~Chen, X.~Xiao, and F.~Alleva, ``Recognizing
  overlapped speech in meetings: A multichannel separation approach using
  neural networks,'' in \emph{InterSpeech}, 2018.

\bibitem{Chen2021ContinuousSS}
S.~Chen, Y.~Wu, Z.~Chen, J.~Li, C.~Wang, S.~Liu, and M.~Zhou, ``Continuous
  speech separation with conformer,'' in \emph{IEEE ICASSP}, 2021.

\bibitem{Wang2022LeveragingRC}
X.~Wang, D.~Wang, N.~Kanda, S.~E. Eskimez, and T.~Yoshioka, ``Leveraging real
  conversational data for multi-channel continuous speech separation,'' in
  \emph{InterSpeech}, 2022.

\bibitem{Chen2022SpeechSW}
Z.~Chen, N.~Kanda, J.~Wu, Y.~Wu, X.~Wang, T.~Yoshioka, J.~Li, S.~Sivasankaran,
  and S.~E. Eskimez, ``Speech separation with large-scale self-supervised
  learning,'' in \emph{IEEE ICASSP}, 2022.

\bibitem{Kinoshita2018ListeningTE}
K.~Kinoshita, L.~Drude, M.~Delcroix, and T.~Nakatani, ``Listening to each
  speaker one by one with recurrent selective hearing networks,'' in \emph{IEEE
  ICASSP}, 2018.

\bibitem{Zhang2021ContinuousSS}
Y.~Zhang, Z.~Chen, J.~Wu, T.~Yoshioka, P.~Wang, Z.~Meng, and J.~Li,
  ``Continuous speech separation with recurrent selective attention network,''
  in \emph{IEEE ICASSP}, 2021.

\bibitem{vonNeumann2021GraphPITGP}
T.~von Neumann, K.~Kinoshita, C.~Boeddeker, M.~Delcroix, and R.~Haeb-Umbach,
  ``{Graph-PIT}: Generalized permutation invariant training for continuous
  separation of arbitrary numbers of speakers,'' in \emph{InterSpeech}, 2021.

\bibitem{vonNeumann2023SegmentLessCS}
------, ``Segment-less continuous speech separation of meetings: Training and
  evaluation criteria,'' \emph{IEEE/ACM TASLP}, vol.~31, pp. 576--589, 2023.

\bibitem{Wang2020MultimicrophoneCS}
Z.-Q. Wang, P.~Wang, and D.~Wang, ``Multi-microphone complex spectral mapping
  for utterance-wise and continuous speech separation,'' \emph{IEEE/ACM TASLP},
  vol.~29, pp. 2001--2014, 2020.

\bibitem{Yoshioka2019LowlatencySC}
T.~Yoshioka, Z.~Chen, C.~Liu, X.~Xiao, H.~Erdogan, and D.~Dimitriadis,
  ``Low-latency speaker-independent continuous speech separation,'' in
  \emph{IEEE ICASSP}, 2019.

\bibitem{Wang2021LocalizationBS}
Z.-Q. Wang and D.~Wang, ``Localization based sequential grouping for continuous
  speech separation,'' in \emph{IEEE ICASSP}, 2021.

\bibitem{barker2018fifth}
J.~Barker, S.~Watanabe, E.~Vincent, and J.~Trmal, ``The fifth {'CHiME'} speech
  separation and recognition challenge: dataset, task and baselines,'' in
  \emph{InterSpeech}, 2018.

\bibitem{Boeddeker2018FrontendPF}
C.~Boeddeker, J.~Heitkaemper, J.~Schmalenstroeer, L.~Drude, J.~Heymann, and
  R.~Haeb-Umbach, ``Front-end processing for the {CHiME-5} dinner party
  scenario,'' in \emph{CHiME Workshop}, 2018.

\bibitem{Horiguchi2020BlockOnlineGS}
S.~Horiguchi, Y.~Fujita, and K.~Nagamatsu, ``Block-online guided source
  separation,'' in \emph{IEEE SLT}, 2021.

\bibitem{Raj2022GPUacceleratedGS}
D.~Raj, D.~Povey, and S.~Khudanpur, ``{GPU}-accelerated guided source
  separation for meeting transcription,'' in \emph{InterSpeech}, 2023.

\bibitem{Kanda2019GuidedSS}
N.~Kanda, C.~B{\"o}ddeker, J.~Heitkaemper, Y.~Fujita, S.~Horiguchi,
  K.~Nagamatsu, and R.~H{\"a}b-Umbach, ``Guided source separation meets a
  strong {ASR} backend: Hitachi/paderborn university joint investigation for
  dinner party {ASR},'' in \emph{InterSpeech}, 2019.

\bibitem{Arora2020TheJM}
A.~Arora, D.~Raj, A.~S. Subramanian, K.~Li, B.~Ben-Yair, M.~Maciejewski,
  P.~Żelasko, L.~P. Garc{\'i}a-Perera, S.~Watanabe, and S.~Khudanpur, ``The
  {JHU} multi-microphone multi-speaker {ASR} system for the {CHiME-6}
  challenge,'' in \emph{The CHiME Workshop}, 2020.

\bibitem{Medennikov2020TheSS}
I.~Medennikov, M.~Korenevsky, T.~Prisyach, Y.~Y. Khokhlov, M.~Korenevskaya,
  I.~Sorokin, T.~Timofeeva, A.~Mitrofanov, A.~Andrusenko, I.~Podluzhny,
  A.~Laptev, and A.~Romanenko, ``The {STC} system for the {CHiME-6}
  challenge,'' in \emph{CHiME Workshop}, 2020.

\bibitem{Boeddeker2022AnIS}
C.~Boeddeker, T.~Cord-Landwehr, T.~von Neumann, and R.~Haeb-Umbach, ``An
  initialization scheme for meeting separation with spatial mixture models,''
  in \emph{InterSpeech}, 2022.

\bibitem{molkov2021AuxiliaryLF}
K.~Žmol{\'i}kov{\'a}, M.~Delcroix, D.~Raj, S.~Watanabe, and J.~H.
  Cernock{\'y}, ``Auxiliary loss function for target speech extraction and
  recognition with weak supervision based on speaker characteristics,'' in
  \emph{InterSpeech}, 2021.

\bibitem{Sivaraman2021AdaptingSS}
A.~Sivaraman, S.~Wisdom, H.~Erdogan, and J.~R. Hershey, ``Adapting speech
  separation to real-world meetings using mixture invariant training,'' in
  \emph{IEEE ICASSP}, 2021.

\bibitem{Yu2022MFCCAMultiFrameCA}
F.~Yu, S.~Zhang, P.~Guo, Y.~Liang, Z.~Du, Y.~Lin, and L.~Xie, ``{MFCCA}:
  Multi-frame cross-channel attention for multi-speaker {ASR} in multi-party
  meeting scenario,'' in \emph{IEEE SLT}, 2022.

\bibitem{Janin2003TheIM}
A.~L. Janin, D.~Baron, J.~Edwards, D.~P.~W. Ellis, D.~Gelbart, N.~Morgan,
  B.~Peskin, T.~Pfau, E.~Shriberg, A.~Stolcke, and C.~Wooters, ``The {ICSI}
  meeting corpus,'' in \emph{IEEE ICASSP}, 2003.

\bibitem{Fox2013TheSW}
C.~W. Fox, Y.~Liu, E.~Zwyssig, and T.~Hain, ``The sheffield wargames corpus,''
  in \emph{InterSpeech}, 2013.

\bibitem{Fu2021AISHELL4AO}
Y.~Fu, L.~Cheng, S.~Lv, Y.~Jv, Y.~Kong, Z.~Chen, Y.~Hu, L.~Xie, J.~Wu, H.~Bu,
  X.~Xu, J.~Du, and J.~Chen, ``{AISHELL-4}: An open source dataset for speech
  enhancement, separation, recognition and speaker diarization in conference
  scenario,'' in \emph{InterSpeech}, 2021.

\bibitem{Yang2022OpenSM}
Z.~Yang, Y.~Chen, L.~Luo, R.~Yang, L.~Ye, G.~Cheng, J.~Xu, Y.~Jin, Q.~Zhang,
  P.~Zhang, L.~Xie, and Y.~Yan, ``Open source magicdata-{RAMC}: A rich
  annotated mandarin conversational speech dataset,'' in \emph{InterSpeech},
  2022.

\bibitem{Sak2017RecurrentNA}
H.~Sak, M.~Shannon, K.~Rao, and F.~Beaufays, ``Recurrent neural aligner: An
  encoder-decoder neural network model for sequence to sequence mapping,'' in
  \emph{InterSpeech}, 2017.

\bibitem{Tripathi2019MonotonicRN}
A.~Tripathi, H.~Lu, H.~Sak, and H.~Soltau, ``Monotonic recurrent neural network
  transducer and decoding strategies,'' in \emph{IEEE ASRU}, 2019.

\bibitem{Moritz2022AnIO}
N.~Moritz, F.~Seide, D.~Le, J.~Mahadeokar, and C.~Fuegen, ``An investigation of
  monotonic transducers for large-scale automatic speech recognition,'' in
  \emph{IEEE SLT}, 2022.

\bibitem{Mahadeokar2020AlignmentRS}
J.~Mahadeokar, Y.~Shangguan, D.~Le, G.~Keren, H.~Su, T.~Le, C.~feng Yeh,
  C.~Fuegen, and M.~L. Seltzer, ``Alignment restricted streaming recurrent
  neural network transducer,'' in \emph{IEEE SLT}, 2021.

\bibitem{Kuang2022PrunedRF}
F.~Kuang, L.~Guo, W.~Kang, L.~Lin, M.~Luo, Z.~Yao, and D.~Povey, ``Pruned
  {RNN-T} for fast, memory-efficient {ASR} training,'' in \emph{InterSpeech},
  2022.

\bibitem{Luo2019DualPathRE}
Y.~Luo, Z.~Chen, and T.~Yoshioka, ``Dual-path {RNN}: Efficient long sequence
  modeling for time-domain single-channel speech separation,'' in \emph{IEEE
  ICASSP}, 2019.

\bibitem{zipformer}
D.~Povey,
  \url{https://github.com/k2-fsa/icefall/blob/master/egs/librispeech/ASR/pruned_transducer_stateless7/zipformer.py}.

\bibitem{Ghodsi2020RnnTransducerWS}
M.~R. Ghodsi, X.~Liu, J.~A. Apfel, R.~Cabrera, and E.~Weinstein,
  ``{RNN-Transducer} with stateless prediction network,'' in \emph{IEEE
  ICASSP}, 2020.

\bibitem{Kim2016JointCB}
S.~Kim, T.~Hori, and S.~Watanabe, ``Joint {CTC}-attention based end-to-end
  speech recognition using multi-task learning,'' in \emph{IEEE ICASSP}, 2016.

\bibitem{Sudo20224DAJ}
Y.~Sudo, M.~H. Shakeel, B.~Yan, J.~Shi, and S.~Watanabe, ``{4D ASR}: Joint
  modeling of {CTC}, attention, transducer, and mask-predict decoders,'' in
  \emph{InterSpeech}, 2023.

\bibitem{Landini2022FromSM}
F.~Landini, A.~Lozano-Diez, M.~D{\'i}ez, and L.~Burget, ``From simulated
  mixtures to simulated conversations as training data for end-to-end neural
  diarization,'' in \emph{InterSpeech}, 2022.

\bibitem{Panayotov2015LibrispeechAA}
V.~Panayotov, G.~Chen, D.~Povey, and S.~Khudanpur, ``Librispeech: An {ASR}
  corpus based on public domain audio books,'' in \emph{IEEE ICASSP}, 2015.

\bibitem{Mir2007AcousticBF}
X.~A. Mir{\'o}, C.~Wooters, and J.~Hernando, ``Acoustic beamforming for speaker
  diarization of meetings,'' \emph{IEEE TASLP}, vol.~15, pp. 2011--2022, 2007.

\bibitem{Renals2014NeuralNF}
S.~Renals and P.~Swietojanski, ``Neural networks for distant speech
  recognition,'' in \emph{4th Joint Workshop on Hands-free Speech Communication
  and Microphone Arrays (HSCMA)}, 2014.

\bibitem{vonNeumann2022OnWE}
T.~von Neumann, C.~Boeddeker, K.~Kinoshita, M.~Delcroix, and R.~Haeb-Umbach,
  ``On word error rate definitions and their efficient computation for
  multi-speaker speech recognition systems,'' in \emph{IEEE ICASSP}, 2023.

\bibitem{Huang2022AdaptingSM}
Z.~Huang, D.~Raj, L.~P. Garc{\'i}a-Perera, and S.~Khudanpur, ``Adapting
  self-supervised models to multi-talker speech recognition using speaker
  embeddings,'' in \emph{IEEE ICASSP}, 2023.

\bibitem{steinmetz2021pyloudnorm}
C.~J. Steinmetz and J.~D. Reiss, ``pyloudnorm: {A} simple yet flexible loudness
  meter in python,'' in \emph{150th AES Convention}, 2021.

\bibitem{Snyder2015MUSANAM}
D.~Snyder, G.~Chen, and D.~Povey, ``{MUSAN}: A music, speech, and noise
  corpus,'' \emph{ArXiv}, 2015.

\bibitem{Park2019SpecAugmentAS}
D.~S. Park, W.~Chan, Y.~Zhang, C.-C. Chiu, B.~Zoph, E.~D. Cubuk, and Q.~V. Le,
  ``Specaugment: A simple data augmentation method for automatic speech
  recognition,'' in \emph{InterSpeech}, 2019.

\bibitem{Boeddeker2023TSSEPJD}
C.~Boeddeker, A.~S. Subramanian, G.~Wichern, R.~Haeb-Umbach, and J.~L. Roux,
  ``{TS-SEP}: Joint diarization and separation conditioned on estimated speaker
  embeddings,'' \emph{ArXiv}, vol. abs/2303.03849, 2023.

\bibitem{Kocour2021RevisitingJD}
M.~Kocour, K.~Žmol{\'i}kov{\'a}, L.~Ondel, J.~Svec, M.~Delcroix, T.~Ochiai,
  L.~Burget, and J.~H. Cernock{\'y}, ``Revisiting joint decoding based
  multi-talker speech recognition with {DNN} acoustic model,'' in
  \emph{InterSpeech}, 2021.

\bibitem{Xia2021TurntoDiarizeOS}
W.~Xia, H.~Lu, Q.~Wang, A.~Tripathi, I.~Lopez-Moreno, and H.~Sak,
  ``Turn-to-diarize: Online speaker diarization constrained by transformer
  transducer speaker turn detection,'' in \emph{IEEE ICASSP}, 2021.

\bibitem{Xu2022RescoreBERTDS}
L.~Xu, Y.~Gu, J.~Kolehmainen, H.~Khan, A.~Gandhe, A.~Rastrow, A.~Stolcke, and
  I.~Bulyko, ``{RescoreBERT}: Discriminative speech recognition rescoring with
  {BERT},'' in \emph{IEEE ICASSP}, 2022.

\end{thebibliography}

\vfill

\end{document}